\documentclass[secthm]{elsart}

\journal{\tocs}

\usepackage[english]{babel}
\usepackage[T1]{fontenc}
\usepackage[latin1]{inputenc}

\usepackage{mathrsfs}
\usepackage{amsfonts}
\usepackage{amssymb}
\usepackage{amsmath}

\usepackage{comment}

\usepackage{pstricks}

\usepackage{color}
\usepackage{graphicx}

\DeclareMathAlphabet{\mathpzc}{OT1}{pzc}{m}{it}

\newcommand{\macro}[3]{\newcommand{#1}[#3]{#2}}

\newcommand\macros[4]
                   {%
                     \newenvironment{#1}[1][#2]{\par\vspace{1ex}
                       #3 \hspace{0.5em}#4}
                                    {\nopagebreak%
                                      \strut\nopagebreak%
                                      \par\vspace{2ex}
                                    }
                   }

\macro{\cwd}{\operatorname{cwd}(#1)}{1}
\macro{\twd}{\operatorname{twd}(#1)}{1}
\macro{\brwd}{\operatorname{brwd}(#1)}{1}
\macro{\rwd}{\operatorname{rwd}(#1)}{1}
\macro{\Frwd}{{#1}\textrm{-}\operatorname{rwd}(#2)}{2}
\macro{\Qrwd}{\Frwd{\gfq}{#1}}{1}
\macro{\frwd}{\Frwd{\bF}{#1}}{1}
\macro{\bwd}{\operatorname{bwd}(#1)}{1}
\macro{\Fbrwd}{{#1}\textrm{-}\operatorname{brwd}(#2)}{2}
\macro{\fbrwd}{\Fbrwd{\bF}{#1}}{1}

\macro{\const}{\mathbf{#1}}{1}
\macro{\angl}{\mathop\langle #1 \mathop\rangle}{1}
\macro{\up}{\ulcorner #1\urcorner}{1}
\macro{\card}{\left|{#1}\right|}{1}
\macro{\floor}{\left\lfloor{#1}\right\rfloor}{1}
\macro{\ceil}{\left\lceil{#1}\right\rceil}{1}
\macro{\pare}{\left({#1}\right)}{1}
\macro{\crochet}{\left[{#1}\right]}{1}
\macro{\set}{\left\{{#1}\right\}}{1}
\macro{\range}{\set{{#1},\ldots,{#2}}}{2}
\macro{\mat}{M_{#1}}{1}
\macro{\matind}{{#1}[{#2},{#3}]}{3}
\macro{\matgind}{\matind{\matg}{#1}{#2}}{2}
\macro{\leaves}{\operatorname{L}_{#1}}{1}
\macro{\cutrk}{{#1}\textrm{-}\operatorname{cutrk}}{1}
\macro{\comp}{(X^{#1},#2\backslash X^{#1})}{2}
\macro{\supp}{\mathpzc{u}(#1)}{1}
\macro{\field}{\mathbb{F}_{#1}}{1}
\macro{\email}{\texttt{#1}}{1}
\macro{\subg}{#1\textrm{-}#2}{2}
\macro{\bicutrk}{{#1}\textrm{-}\operatorname{bicutrk}}{1}

\def\gfa{\mathbb{a}}
\def\gfb{{\gfa^2}}
\def\gfq{\operatorname{\field{4}}}
\def\rk{\operatorname{rk}}

\def\matg{\mat{G}}
\def\fcutrk{\cutrk{\bF}}
\def\fbicutrk{\bicutrk{\bF}}
\def\ucutrk{\operatorname{cutrk}}

\def\ie{\emph{i.e.}}
\def\eg{\emph{e.g.}}

\def\bR{\mathbb{R}}
\def\bN{\mathbb{N}}

\def\bC{\mathbb{C}}

\def\bF{\mathbb{F}}
\def\bG{\mathbb{G}}

\def\cA{\mathcal{A}}

\def\cF{\mathcal{F}}
\def\cP{\mathcal{P}}

\def\cC{\mathcal{C}}

\def\cL{\mathcal{L}}

\def\cR{\mathcal{R}}

\def\cI{\mathcal{I}}
\def\cM{\mathcal{M}}

\def\cBR{\mathcal{BR}}
\def\cBC{\mathcal{BC}}

\def\sG{\mathscr{G}}
\def\sC{\mathscr{C}}
\def\sS{\mathscr{S}}

\def\tG{\widetilde{G}}

\def\hlineny{Hlin$\check{\textrm{e}}$n\'y~}

\def\dam{Discrete Applied Mathematics}

\def\jctb{Journal of Combinatorial Theory, Series B}

\def\jcss{Journal of Computer and System Sciences}
\def\siamjadm{SIAM Journal on Algebraic and Discrete Methods}
\def\siamjc{SIAM Journal on Computing}

\def\endm{Electronic Notes in Discrete Mathematics}
\def\jal{Journal of Applied Logic}
\def\tocs{Theory of Computing Systems}
\def\daor{Discrete Analysis and Operations Research}

\def\stoc{Symposium on Theory of Computing}

\def\ic{Information and Computation}
\def\jgt{Journal of Graph Theory}
\def\mscs{Mathematical Structures in Computer Science}
\def\lap{Linear Algebra and its Applications}
\def\ejc{European Journal of Combinatorics}

\macros{defi}{Example}{\noindent{\bf #1}}{}
\macros{mthm}{Main Theorem}{\noindent{\bf #1}}{\itshape}

\newcounter{CaseCtr}
               {\end{list}}

\usepackage{calc}

{\end{list}}

\def\lc#1#2{{#1}*{(#2,\lambda)}}
\def\lco#1#2{{#1}*{#2}}
\def\pc#1#2{{#1}\wedge{#2}}

\def\colgr{\sG}
\def\fieldgr{\sS}

\begin{document}

\begin{frontmatter}

\title{The Rank-Width of Edge-Colored Graphs}

\author[KAN]{Mamadou Moustapha Kant\'e\thanksref{ANR}} \and
\author[RAO]{Michael Rao\thanksref{ANR}}

\address[KAN]{Clermont-Université, Université Blaise Pascal, LIMOS,
  CNRS\\Complexe Scientifique des Cézeaux 63173 Aubiére Cedex,
  France\\ \email{mamadou.kante@isima.fr}}

\address[RAO]{{Université Bordeaux 1, LaBRI, CNRS}\\ 351 Cours de
la lib\'eration 33405 Talence Cedex, France. \\
\email{rao@labri.fr}.}  

\thanks[ANR]{A part of this research was supported by the project "Graph decompositions
  and algorithms (GRAAL)'' of ``Agence Nationale Pour la Recherche''
  and was done when the first author was in Université Bordeaux 1, LaBRI.}

\begin{abstract} \emph{Clique-width} is a complexity measure of
  directed as well as undirected graphs. \emph{Rank-width} is an
  equivalent complexity measure for undirected graphs and has good
  algorithmic and structural properties. It is in particular related
  to the \emph{vertex-minor} relation. We discuss an extension of the
  notion of rank-width to edge-colored graphs. A $C$-colored graph is
  a graph where the arcs are colored with colors from the set
  $C$. There is not a natural notion of rank-width
  for $C$-colored graphs. We define two notions of rank-width for
  them, both based on a coding of $C$-colored graphs by edge-colored
  graphs where each edge has exactly one color from a field 
  $\bF$ and named respectively \emph{$\bF$-rank-width} and
  \emph{$\bF$-bi-rank-width}. The two notions are equivalent to
  clique-width.  We then present a notion of \emph{vertex-minor} for
  $\bF$-colored graphs and prove that $\bF$-colored graphs of bounded
  $\bF$-rank-width are characterised by a finite list of $\bF$-colored
  graphs to exclude as vertex-minors. A cubic-time algorithm to decide
  whether a $\bF$-colored graph has $\bF$-rank-width
  (resp. $\bF$-bi-rank-width) at most $k$, for fixed $k$, is also
  given. Graph operations to check MSOL-definable properties on
  $\bF$-colored graphs of bounded rank-width are presented. A
  specialisation of all these notions to (directed) graphs without
  edge colors is presented, which shows that our results
  generalise the ones in undirected graphs.
\end{abstract}


\end{frontmatter}

\section{Introduction}\label{sec:1}

\emph{Clique-width} \cite{CER93,CO00} is a complexity measure for
edge-colored graphs, \ie, graphs where edges are colored with colors
from a finite set. Clique-Width is more general than \emph{tree-width}
\cite{RS86} because every class of graphs of bounded tree-width has
bounded clique-width and the converse is false (complete undirected
graphs have clique-width $2$ and unbounded tree-width)
\cite{CO00}. Clique-width is an interesting complexity measure in
algorithmic design.  In fact every property expressible in
\emph{monadic second-order logic} (MSOL for short)
can be checked in linear-time, provided the clique-width expression is
given, on every graph that has small clique-width \cite{CMR00}. This
result is important in complexity theory because many NP-complete
problems are MS-definable properties, \eg, $3$-colorability. However,
it is NP-complete to check if a graph has clique-width at most $k$
when $k$ is part of the input \cite{FRRS06}. It is still open whether
this problem is polynomial for fixed $k\geq 4$.

In their investigations of a recognition algorithm for undirected
graphs of clique-width at most $k$, for fixed $k$, Oum and Seymour
\cite{OS06} introduced the notion of \emph{rank-decomposition} and
associated complexity measure \emph{rank-width}, of undirected
graphs. Rank-width is defined in a combinatorial way and is equivalent
to the clique-width of undirected graphs in the sense that a class of
graphs has bounded clique-width if and only if it has bounded
rank-width \cite{OS06}. But, being defined in a combinatorial way
provides to rank-width better algorithmic properties than
clique-width, in particular:
\begin{itemize}
\item for fixed $k$, there exists a cubic-time algorithm that decides
  whether the rank-width of an undirected graph is at most $k$ and if
  so, constructs a rank-decomposition of width at most $k$
  \cite{HO07};

\item there exists a dual notion to rank-width, the notion of
  \emph{tangle} \cite{LMT10,RS91}. This dual notion is
  interesting for getting certificates in recognition algorithms. 
\end{itemize}

Since clique-width and rank-width of undirected graphs are equivalent,
one way to check MSOL properties in undirected graphs of small
rank-width is to transform a rank-decomposition into a clique-width
expression \cite{OS06}. However, an alternative characterization of
rank-width in terms of graph operations has been proposed in
\cite{CK09}. It is thus possible to solve MSOL properties in
graphs of small rank-width by using directly the
rank-decomposition. This later result is important in a practical
point of view because it avoids the exponent, that cannot be avoided
\cite{CR05,OUM08}, when transforming a rank-decomposition into a
clique-width expression.

Another advantage of rank-width over clique-width is that it is
invariant with respect to the \emph{vertex-minor} relation (no such
notion, except for induced sub-graph relation, is known for
clique-width), \ie, if $H$ is a vertex-minor of $G$, then the
rank-width of $H$ is at most the rank-width of $G$ \cite{OUM05}.
Moreover, every class of undirected graphs of bounded rank-width is
characterised by a finite list of undirected graphs to exclude as
vertex-minors \cite{OUM05}. This later result generalises the one of
Robertson and Seymour on undirected graphs of bounded tree-width
\cite{RS86}.

Despite all these positive results of rank-width, the fact that
clique-width is defined for graphs - directed or not, with edge colors
or not - is an undeniable advantage over rank-width. It is thus
natural to ask for a notion of rank-width for edge-colored graphs or
at least for directed graphs without edge colors. Courcelle and Oum
suggested in \cite{CO07} a definition of rank-width for directed
graphs as follows: Courcelle \cite{COU06} described a graph
transformation $B$ from (directed) graphs to undirected bipartite
graphs so that $f_1(\cwd{B(G)})\leq \cwd{G} \leq f_2(\cwd{B(G)})$, for
some functions $f_1$ and $f_2$; the rank-width of a (directed) graph
is defined as the rank-width of $B(G)$. This definition can be
extended to edge-colored graphs by using a similar coding (see
\cite[Chapter 6]{COU10}). This definition gives a cubic-time algorithm
that approximates the clique-width of edge-colored graphs. Another
consequence is the proof of a weak version of the Seese's conjecture
for edge-colored graphs \cite{CO07}.  However, this definition suffers
from the following drawback: a vertex-minor of $B(G)$ does not always
correspond to a coding of an edge-colored graph and similarly for the
notion of \emph{pivot-minor} (see for instance \cite{FON96,OUM05} for
the definition of pivot-minor of undirected graphs).

We investigate in this paper a better notion of rank-width for
edge-colored graphs. However, there is no unique natural way to extend
rank-width to edge-colored graphs. We are looking for a notion that
extends the one on undirected graphs and that can be used for directed
graphs without edge colors.  For that purposes, we will define the
notion of \emph{sigma-symmetric} matrices, which generalizes the
notion of symmetric and skew-symmetric matrices. We then use this
notion to represent edge-colored graphs by matrices over finite fields
and derive, from this representation, a notion of rank-width, called
\emph{$\bF$-rank-width}, that generalises the one of undirected
graphs. We also define another notion of rank-width, called
\emph{$\bF$-bi-rank-width}. We prove that the two parameters are
equivalent to clique-width.

We then define a notion of vertex-minor for edge-colored graphs that
extends the one on undirected graphs. We prove that $\bF$-rank-width
and $\bF$-bi-rank-width are invariant with respect to this
vertex-minor relation. We give a characterisation of edge-colored
graphs of bounded $\bF$-rank-width by excluded configurations. This
result generalises the one on undirected graphs \cite{OUM05}. A
generalisation of the pivot-minor relation is also presented. 

The cubic-time recognition algorithm by \hlineny and Oum \cite{HO07}
can be adapted to give for fixed $k$, a cubic-time algorithm that
decides if a given edge-colored graph has $\bF$-rank-width
(resp. $\bF$-bi-rank-width) at most $k$ and if so, outputs an optimal
rank-decomposition.

The two notions of rank-width of edge-colored graphs are specialised
to directed graphs without colors on edges. All the results
specialised to them. 

The paper is organized as follows. In Section \ref{sec:2} we give some
preliminary definitions and results. We recall in particular the
definition of rank-width of undirected graphs. The first notion of
rank-width of edge-colored graphs, called $\bF$-rank-width, is studied
in Section \ref{sec:3}. We will define the notion of vertex-minor and
pivot-minor, and prove that edge-colored graphs of bounded
$\bF$-rank-width are characterised by a finite list of edge-colored
graphs to exclude as vertex-minors (resp. pivot-minors). A cubic-time
recognition algorithm and a specialisation to directed graphs are also
presented.  We define our second notion of rank-width for edge-colored
graphs called $\bF$-bi-rank-width in Section \ref{sec:4}. We also
specialise it to directed graphs. In Section \ref{sec:5} we introduce
some algebraic graph operations that generalise the ones in
\cite{CK09}. These operations will be used to characterise exactly the
two notions of rank-width. They can be seen as alternatives to
clique-width operations for solving MSOL properties. We conclude by
some remarks and open questions in Section \ref{sec:6}.

This paper is related to a companion paper where the authors introduce
a decomposition of edge-colored graphs on a fixed
field~\cite{KR10b}. This decomposition plays a role similar to the
\emph{split decomposition} for the rank-width of undirected
graphs. Particularly we show that the rank width of an edge-colored
graph is exactly the maximum over the rank-width over all edge-colored
prime graphs in the decomposition, and we give different
characterisations of egde-colored graphs of rank-width one.

\section{Preliminaries}\label{sec:2}

For two sets $A$ and $B$, we let $A\backslash B$ be the set $\{x\in
A\mid x\notin B\}$. The power-set of a set $V$ is denoted by $2^V$. We
often write $x$ to denote the set $\{x\}$. The set of natural integers
is denoted by $\bN$. 

We denote by $+$ and $\cdot$ the binary operations of any field and by
$0$ and $1$ the neutral elements of $+$ and $\cdot$ respectively. For
every prime number $p$ and every positive integer $k$, we denote by
$\field{p^k}$ the finite field of characteristic $p$ and of order $p^k$. 
We recall that they are the only finite fields. We refer to \cite{LN97} 
for our field terminology. 

For sets $R$ and $C$, an \emph{$(R,C)$-matrix} is a matrix where the
rows are indexed by elements in $R$ and columns indexed by elements in
$C$. For an $(R,C)$-matrix $M$, if $X\subseteq R$ and $Y\subseteq C$,
we let $\matind{M}{X}{Y}$ be the sub-matrix of $M$ where the rows and
the columns are indexed by $X$ and $Y$ respectively. We let $\rk$ be
the matrix rank-function (the field will be clear from the
context). We denote by $M^T$ the transpose of a matrix $M$. The order
of an $(R,C)$-matrix is defined as $|R|\times |C|$. We often write
$k\times \ell$-matrix to denote a matrix of order $k\times \ell$.
For positive integers $k$ and $\ell$, we let $O_{k,\ell}$ be the null
$k\times \ell$-matrix and $I_k$ the identity $k\times k$-matrix, or
respectively $O$ and $I$ when the size is clear in the context.

We use the standard graph terminology, see for instance
\cite{DIE05}. A graph $G$ is a couple $(V_G,E_G)$ where $V_G$ is the
set of vertices and $E_G\subseteq V_G\times V_G$ is the set of
edges. A graph $G$ is said to be \emph{oriented} if $(x,y)\in E_G$
implies $(y,x)\notin E_G$, and it is said \emph{undirected} if
$(x,y)\in E_G$ implies $(y,x)\in E_G$.  An edge between $x$ and $y$ in
an undirected graph is denoted by $xy$ (equivalently $yx$). For a
graph $G$, we denote by $G[X]$, called the sub-graph of $G$ induced by
$X\subseteq V_G$, the graph $(X,E_G\cap (X\times X))$; we let
$\subg{G}{X}$ be the sub-graph $G[V_G\backslash X]$.  The degree of a
vertex $x$ in an undirected graph $G$ is the cardinal of the set
$\{y\mid xy\in E_G\}$.  Two graphs $G$ and $H$ are \emph{isomorphic}
if there exists a bijection $h:V_G\to V_H$ such that $(x,y)\in E_G$ if
and only if $(h(x),h(y))\in E_H$. We call $h$ an \emph{isomorphism}
between $G$ and $H$. All graphs are finite and loop-free (\emph{i.e.}
for every $x\in V_G$, $(x,x)\not\in E_G$).

A \emph{tree} is an acyclic connected undirected graph. In order to
avoid confusions in some lemmas, we will call \emph{nodes} the
vertices of trees. The nodes of degree $1$ are called \emph{leaves}
and the set of leaves in a tree $T$ is denoted by $\leaves{T}$. A
\emph{sub-cubic tree} is an undirected tree such that the degree of
each node is at most $3$.  A tree $T$ is \emph{rooted} if it has a
distinguished node $r$, called the \emph{root} of $T$. For
convenience, we will consider a rooted tree as an oriented graph such
that the underlying graph is a tree, and such that all nodes are
reachable from the root by a directed path.  For a tree $T$ and an
edge $e$ of $T$, we let $\subg{T}{e}$ denote the graph
$(V_T,E_T\backslash e)$.

Let $C$ be a (possibly infinite) set that we call the \emph{colors}.
A \emph{$C$-graph} $G$ is a tuple $(V_G, E_G, \ell_G)$ where
$(V_G,E_G)$ is a graph and $\ell_G: E_G \to C$ is a function. Its
associated \emph{underlying graph} $\supp{G}$ is the graph
$(V_G,E_G)$. Two $C$-graph $G$ and $H$ are isomorphic if there is a
isomorphism $h$ between $(V_G, E_G)$ and $(V_H,E_H)$ such that for
every $(x,y)\in E_G$, $\ell_G((x,y))=\ell_H((h(x),h(y))$.  We call
$h$ an \emph{isomorphism} between $G$ and $H$.  We let $\colgr(C)$ be
the set of $C$-graphs for a fixed color set $C$.  Even though we
authorise infinite color sets in the definition, most of results in
this article are valid only when the color set is finite.  It is worth
noticing that an edge-uncolored graph can be seen as an edge-colored
graph where all the edges have the same color.

\begin{rem}[Multiple colors per edge]\label{rem:1.1} In our
  definition, an edge in a $C$-graph can only have one color. However,
  this is not restrictive because if in an edge-colored graph an edge
  can have several colors from a set $C$, we just extend $C$ to $2^C$.
\end{rem}

\begin{rem}[$2$-structures and edge-colored graphs]\label{rem:1.2} A
  \emph{$2$-structure} \cite{EHR99} is a pair
  $(D,R)$ where $D$ is a finite set and $R$ is an equivalence relation on 
  the set $D_2 = \{(x,y)\mid x,y\in D \text{ and }\ x\ne
  y\}$. Every $2$-structure $(D,R)$ can be seen as a $C$-colored
  graph $G=(D,D_2,\ell)$ where $C:=\{[e]\mid [e]$ is an equivalence class of
  $R\}$ and for every edge $e,\ \ell(e):=[e]$. Equivalently,
  every $C$-graph $G$ can be seen as a $2$-structure $(V_G,R)$ where $eRe'$ 
  if and only if $\ell_G(e)=\ell_G(e')$ and all the non-edges in
  $G$ are equivalent with respect to $R$.
\end{rem}


A \emph{parameter} on $\colgr(C)$ is a function $wd:\colgr(C)\to \bN$ that is
invariant under isomorphism. Two parameters on $\colgr(C)$, say
$wd$ and $wd'$, are \emph{equivalent} if there exist two 
integer functions $f$ and $g$ such that for every
edge-colored graph $G\in \colgr(C)$, $f(wd'(G)) \leq wd(G) \leq
g(wd'(G))$.


The \emph{clique-width}, denoted by $\operatorname{cwd}$, is a graph parameter 
defined by Courcelle et al. \cite{CER93,CO00}. 
Most of the investigations concern edge-uncolored graphs. 
However, its edge-colored version has been
investigated these last years (see \cite{CB06,FMR08}). Note that 
the clique-width is also defined in more general case 
where edges can have several colors.

We finish these preliminaries by the notion of terms. Let $\cF$ be a
set of binary and unary function symbols and $\cC$ a set of
constants. We denote by $T(\cF,\cC)$ the set of finite well-formed
terms built with $\cF\cup \cC$. Notice that the syntactic tree of a
term is rooted.

A \emph{context} is a term in $T(\cF,\cC\cup\{u\})$ having a single
occurrence of the variable $u$ (a nullary symbol). We denote by
$Cxt(\cF,\cC)$ the set of contexts. We denote by $Id$ the particular
context $u$. If $s$ is a context and $t$ a term, we let $s\bullet t$
be the term in $T(\cF,\cC)$ obtained by substituting $t$ for $u$ in $s$. 



\subsection{Rank-Width and Vertex-Minor of Undirected Graphs}\label{subsec:2.2}

Despite the interesting algorithmic results \cite{CMR00}, clique-width
suffers from the lack of a recognition algorithm. In their
investigations for a recognition algorithm, Oum and Seymour introduced
the notion of \emph{rank-width} \cite{OS06}, which approximates the
clique-width of undirected graphs.  Let us first define some notions.

Let $V$ be a finite set and $f:2^V\to \bN$ a function. We say that
$f$ is \emph{symmetric} if for any $X\subseteq V, ~f(X)=f(V\backslash
X)$; $f$ is \emph{submodular} if for any $X,Y\subseteq V$, $f(X\cup Y)
+ f(X\cap Y) \leq f(X) +f(Y)$.

A \emph{layout} of a finite set $V$ is a pair $(T,\cL)$ of a sub-cubic
tree $T$ and a bijective function $\cL:V\to \leaves{T}$. For each edge
$e$ of $T$, the connected components of $\subg{T}{e}$ induce a bipartition
$(X_e,V\backslash X_e)$ of $\leaves{T}$, and thus a bipartition
$\comp{e}{V}= (\cL^{-1}(X_e), \cL^{-1}(V\backslash X_e))$ of $V$ (we
will omit the sub or sup-script $e$ when the context is clear).

Let $f:2^V\to \bN$ be a symmetric function and $(T,\cL)$ a layout of
$V$. The \emph{$f$-width of each edge $e$ of $T$} is defined as
$f(X^e)$ and the \emph{$f$-width of $(T,\cL)$} is the maximum
$f$-width over all edges of $T$. The \emph{$f$-width of $V$} is the
minimum $f$-width over all layouts of $V$.

\begin{defn}[Rank-width of undirected graphs \cite{OUM05,OS06}] For
  every undirected graph $G$, we let $\matg$ be its adjacency
  $(V_G,V_G)$-matrix where $\matgind{x}{y}:=1$ if and only if $xy\in
  E_G$. For every graph $G$, we let $\ucutrk_G:2^{V_G}\to \bN$ where
  $\ucutrk_G(X) := \rk(\matgind{X}{V_G\backslash X})$, where $\rk$ is the matrix rank over $\bF_2$. This function
  is symmetric. The \emph{rank-width} of an undirected graph $G$,
  denoted $\rwd{G}$, is the $\ucutrk_G$-width of $V_G$.
\end{defn}

Rank-Width has several structural and algorithmic results, see for
instance \cite{CK09,HO07,OUM05}. In particular, for fixed $k$, there
exists a cubic-time algorithm for recognizing undirected graphs of
rank-width at most $k$ \cite{HO07}. Moreover, rank-width is related to
a relation on undirected graphs, called \emph{vertex-minor}.

\begin{defn}[Local complementation, Vertex-minor \cite{OUM05}] For an undirected graph $G$
  and a vertex $x$ of $G$, the \emph{local complementation at $x$}, 
  denoted by $G*v$,  consists in replacing the sub-graph induced on the neighbors of $x$
  by its complement. A graph $H$ is a \emph{vertex-minor} of a graph
  $G$ if $H$ can be obtained from $G$ by applying a sequence of local
  complementations and deletions of vertices.
\end{defn}

Authors of \cite{BOU87,FON96,OUM05} also introduced the \emph{pivot}
operation on an edge $xy$, denoted by $\pc{G}{xy}=G*x*y*x = G*y*x*y$.
An interesting theorem relating rank-width and the notion of
vertex-minor is the following.

\begin{thm}[\cite{OUM05}]\label{thm:2.1} For every positive integer
  $k$, there exists a finite list $\sC_k$ of undirected graphs such
  that an undirected graph has rank-width at most $k$ if and only if
  it does not contain as vertex-minor any graph isomorphic to a graph
  in $\sC_k$.
\end{thm}

In the next section, we define the notion of rank-width of
edge-colored graphs and generalize Theorem \ref{thm:2.1} to them.
 
\section{$\bF$-Rank-Width of $\sigma$-Symmetic $\bF^*$-Graphs}\label{sec:3}

We want a notion of rank-width for edge-colored graphs that
generalises the one on undirected graphs. For that purposes, we will
identify each color by an non-zero element of a field. This representation will
allow us to define the rank-width of edge-colored graphs by using rank
matrices.

Let $\bF$ be a field, and let $\bF^*=\bF\setminus \{0\}$ (where $0$ is
the zero of $\bF$).  One can note that there is a natural bijection
between the class of $\bF^*$-graphs and the class of $\bF$-graphs with
complete underlying graph (replace every non-edge by an edge of color
$0$).  From now on, we do not distinguish these two classes,
and we let $\ell_G((x,y))=0$ for all $(x,y)\notin E_G$.

We can represent every $\bF^*$-graph $G$ by a $(V_G,V_G)$-matrix
$\matg$ such that $\matgind{x}{y}:=\ell_G((x,y))$ for every $x,y\in
V_G$ with $x\ne y$, and $\matgind{x}{x}:=0$ for every $x\in V_G$.



Let $\sigma:\bF\to \bF$ be a bijection. We recall that $\sigma$ is an
\emph{involution} if $\sigma(\sigma(a)) = a$ for all $a\in \bF$. We
call $\sigma$ a \emph{sesqui-morphism} if $\sigma$ is an involution,
and the mapping $[x\mapsto \sigma(x)/\sigma(1)]$ is an
automorphism. It is worth noticing that if $\sigma:\bF\to \bF$ is a
sesqui-morphism, then $\sigma(0)=0$ and for every $a,b\in \bF$,
$\sigma(a+b)=\sigma(a)+\sigma(b)$ (\emph{i.e.} $\sigma$ is an
automorphism for the addition). Moreover, we have the following
notable equalities.

\begin{prop} If $\sigma$ is a sesqui-morphism, then
\begin{align*} 
  \sigma(a\cdot b) &= \frac{\sigma(a)\cdot \sigma(b)}{\sigma(1)},\\
  \sigma \left (\frac{a}{b} \right)&= \frac{\sigma(1)\cdot \sigma(a)}{\sigma(b)},\\ 
  \sigma \left (\frac{a \cdot b}{c} \right )&= \frac{\sigma(a)\cdot \sigma(b)}{\sigma(c)}.
\end{align*}
\end{prop}

A $\bF^*$-graph is \emph{$\sigma$-symmetric} if the underlying graph 
is undirected, and for every arc $(x,y)$, $\ell_G((x,y))=a$ if and 
only if $\ell_G((y,x))=\sigma(a)$.
Clearly, if $G$ is a $\sigma$-symmetric
$\bF^*$-graph, then $\matgind{x}{y} = \sigma(\matgind{y}{x})$.  We
denote by $\fieldgr(\bF)$ (respectively $\fieldgr(\bF,\sigma)$) the set of $\bF^*$-graphs
(respectively $\sigma$-symmetric $\bF^*$-graphs). 
Note that $\fieldgr(\bF) = \colgr(\bF^*)$.



To represent a $C$-graph, one can take an injection from $C$ to $\bF^*$
for a large enough field $\bF$. Notice that the representation is not
unique: on one hand, several incomparable fields are possible for
$\bF$, and on the other hand, the representation depends on the
injection from $C$ to $\bF^*$. For example, oriented graphs can be
represented by a $\bF_3^*$-graph or by a $\bF_4^*$-graph (see
Section~\ref{subsec:3.4}). Two different representations can give two
different rank-width parameters, but the two parameters are equivalent
when $C$ is finite (direct consequence of Proposition~\ref{prop:3.4}).


Let $\bF$ be a finite field of characteristic $p$ and order $q$. We will prove that every $\bF^*$-graph
can be seen as a $\widetilde{\sigma}$-symmetric $(\bF^2)^*$-graph for
some sesqui-morphism $\widetilde{\sigma}$, where $\bF^2$ is an algebraic 
extension of $\bF$ of order $2$. Let us first make some observations.

\begin{lem}\label{claim:3.1} There exists an element $p$ in $\bF^*$
  such that the polynomial $X^2 - p(X+1)$ has no root in $\bF$.
\end{lem}

\begin{pf*}{Proof.} There exist
  $|\bF|-1$ distinct polynomials of the form $X^2-p(X+1)$, $p\ne
  0$. We first notice that $0$ or $-1$ cannot be a root of
  $X^2-p(X+1)$, for any $p\in \bF^*$. Now, two such polynomials cannot
  have a common root. Assume the contrary and let $\alpha$ be a root
  of $X^2-p(X+1)$ and of $X^2-p'(X+1)$ with $p\ne p'$. Then
  $(\alpha+1)\cdot (p-p') = 0$, \emph{i.e.} $p=p'$ since $\alpha\ne
  -1$, a contradiction.  Since $-1$ and $0$ cannot be the roots of any
  of the polynomials, we have at most $|\bF|-2$ possible
  roots. Therefore, there exists a $p$ such that $X^2-p(X+1)$ has no
  root in $\bF$.\qed
\end{pf*}

We can now construct an algebraic extension of the finite field $\bF$.
Let $p\in \bF^*$ such that $X^2-p(X+1)$ has no root in $\bF$ and let
$\bF^2$ be isomorphic to the field $\bF[X]\mod (X^2-p(X+1))$
(\emph{i.e.}  $\bF^2$ is the finite field of characteristic $p$ and
order $q^2$).  Let $\alpha:=X\mod (X^2-p(X+1))$. Then every element of
$\bF^2$ is a polynomial on $\alpha$ of the form $a_0+ a_1\alpha$ where
$a_0,a_1\in \bF$. Moreover, $\alpha$ is a root of $X^2-p(X+1)$ in
$\bF^2$.

We let $\gamma:=1-p^{-1}\alpha$ and $\tau:=p^{-1}\alpha$ be in
$\bF^2$. Notice that $\alpha=p\tau$ and $1=\gamma+\tau$.

\begin{lem}\label{claim:3.2} We have the following equalities: 
  \begin{align*}
    \gamma^2 &= (1+p^{-1})\gamma + p^{-1}\tau,\\
    \tau^2 & = p^{-1}\gamma + (1+p^{-1})\tau,\\
    \gamma \cdot \tau &= p^{-1}\gamma + p^{-1}\tau.
  \end{align*}
\end{lem} 


To every pair of elements in $\bF$, we associate an element in
$\bF^2$ by letting $\widetilde{f}:\bF\times \bF\to \bF^2$ where, for
every $(a,b)\in \bF\times \bF$, $\widetilde{f}(a,b):=a\gamma + b\tau$.

\begin{lem}\label{claim:3.3} $\widetilde{f}$ is a
  bijection. 
\end{lem} 


For the sesqui-morphism in $\bF^2$, we let
$\widetilde{\sigma}:\bF^2\to \bF^2$ where $\widetilde{\sigma}(a\gamma
+ b\tau) := b\gamma + a\tau$. One easily verifies that
$\widetilde{\sigma}(\widetilde{\sigma}(\beta)) =\beta$ for all
$\beta\in \bF^2$.
  
\begin{lem}\label{claim:3.4} $\widetilde{\sigma}$ is an
  automorphism. \end{lem}  

\begin{pf*}{Proof.} 
%
  An easy computation shows that $\widetilde{\sigma}((a\gamma +b\tau) + (c\gamma + d\tau)) = \widetilde{\sigma} (a\gamma +b\tau) +  \widetilde{\sigma}(c\gamma +d\tau)$. For the product, we have:
  \begin{align*}
    \widetilde{\sigma}((a\gamma+b\tau)\cdot (c\gamma+d\tau)) &=
    \widetilde{\sigma}(ac\gamma^2 + (ad+bc)\gamma\tau +
    bd\tau^2)\\ &=  ac\widetilde{\sigma}(\gamma^2) +
    (ad+bc)\widetilde{\sigma}(\gamma\tau) +
    bd \widetilde{\sigma}(\tau^2)\\\intertext{and} 
    \widetilde{\sigma}(a\gamma+b\tau) \cdot
    \widetilde{\sigma}(c\gamma+d\tau) &= (b\gamma+a\tau) \cdot
    (d\gamma+c\tau)\\ &= bd\gamma^2 +(ad+bc)\gamma\tau +
    ac\tau^2. 
  \end{align*}
  By Lemma \ref{claim:3.2}, $\widetilde{\sigma}(\gamma^2) = \tau^2$,
  $\widetilde{\sigma}(\tau^2) = \gamma^2$ and
  $\widetilde{\sigma}(\gamma\tau) = \gamma\tau$.  This concludes the proof of the lemma. \qed
\end{pf*}

For every $\bF^*$-graph $G$, we let $\tG$ be the $(\bF^2)^*$-graph
$(V_G,E_G,\ell_{\tG})$ where, for every two distinct vertices $x$ and
$y$,
\begin{align*}
  \ell_{\tG}((x,y)) := \widetilde{f}(\ell_G((x,y)),\ell_G((y,x))).
\end{align*}

By the definitions of $\tG$ and $\widetilde{\sigma}$, and Lemmas
\ref{claim:3.2}-\ref{claim:3.4}, we get the following.

\begin{prop}\label{prop:3.2} 
  The mapping $[G\mapsto \tG]$ from
  $\fieldgr(\bF)$ to $\fieldgr(\bF^2,\widetilde{\sigma})$ is a bijection and for every
  $\bF^*$-graph $G$, $\tG$ is $\widetilde{\sigma}$-symmetric. 
  Moreover, for two $\bF^*$-graphs $G$ and $H$, $\tG$ and $\widetilde{H}$ are
  isomorphic if and only if $G$ and $H$ are isomorphic.
\end{prop}

Nevertheless, two different mappings can give two different rank-width
parameters. But again, since $\bF$ is finite, the parameters are
equivalent.



%

If $\bF$ is infinite, a mapping from $\fieldgr(\bF)$ to
$\fieldgr(\bG,\sigma)$ is not always possible with the previous
construction. For example, a mapping is possible from $\fieldgr(\bR)$
to $\fieldgr(\bC,\sigma)$ with $f(a,b)=(1+i)a+(1-i)b$ and
$\sigma(a+ib)=a-ib$ (where $a,b\in \bR$), but the construction fails
for $\bF=\bC$ since the complexes are algebraically closed.

\medskip

From now on, we will focus our attention to sigma-symmetric
$\bF^*$-graphs.  In Section \ref{subsec:3.1} we define the notion of
\emph{$\bF$-rank-width}. The notion of \emph{vertex-minor} for
$\bF^*$-graphs is presented in Section \ref{subsec:3.2} and we prove
that sigma-symmetric $\bF^*$-graphs of $\bF$-rank-width at most $k$
are characterised by a finite list of sigma-symmetric $\bF^*$-graphs
to exclude as vertex-minors. We prove in Section \ref{subsec:3.3} that
$\bF^*$-graphs of $\bF$-rank-width at most $k$, for fixed $k$, can be
recognised in cubic-time when $\bF$ is finite.  
A specialisation to graphs without colors
on edges is presented in Section \ref{subsec:3.4}.


\subsection{Rank-Width of $\sigma$-symmetric $\bF^*$-Graphs}\label{subsec:3.1}

Along this section, we let $\bF$ be a fixed field (of characteristic
$p$ and of order $q$ if $\bF$ is finite), and we let $\sigma:\bF\to
\bF$ be a fixed sesqui-morphism. We recall that if $G$ is a $\bF^*$-graph, we
denote by $\matg$ the $(V_G,V_G)$-matrix where:
\begin{align*}
  \matgind{x}{y} & := \begin{cases} \ell_G((x,y)) & \textrm{if $x\ne
      y$},\\ 0 & \textrm{otherwise}. \end{cases} 
\end{align*}

\begin{defn}[Cut-Rank Functions]\label{defn:3.1} The
  \emph{$\bF$-cut-rank} function of a $\sigma$-symmetric
  $\bF^*$-graph $G$ is the function $\fcutrk_G:2^{V_G}\to \bN$ where
  $\fcutrk_G(X) = \rk(\matgind{X}{V_G\backslash X})$ for all
  $X\subseteq V_G$.
\end{defn}

\begin{lem}\label{lem:3.1} For every $\sigma$-symmetric
  $\bF^*$-graph $G$, the function $\fcutrk_G$ is symmetric
  and submodular.
\end{lem}

We first recall the \emph{submodular inequality} of the matrix
rank-function.

\begin{prop}{\cite[Proposition 4.1]{OUM05}}\label{prop:3.3} Let $M$
  be an $(R,C)$-matrix over a field $\bF$. Then for all
  $X_1,Y_1\subseteq R$ and $X_2,Y_2\subseteq C$,
  \begin{small}
  \begin{align*}
    \rk(\matind{M}{X_1}{X_2}) + rk(\matind{M}{Y_1}{Y_2})
    &\geq \rk(\matind{M}{X_1\cup Y_1}{X_2\cap Y_2}) +
    \rk(\matind{M}{X_1\cap Y_1}{X_2\cup Y_2}).
  \end{align*}
  \end{small}
\end{prop}

\begin{pf*}{Proof of Lemma \ref{lem:3.1}.} Let $X$ and $Y$ be subsets
  of $V_G$. We let $A_1=\matgind{X}{V_G\backslash X}$ and
  $A_2=\matgind{Y}{V_G\backslash Y}$. We first prove the first
  statement.

  We let $M'$ be the $(V_G\backslash X,X)$-matrix where
  $\matind{M'}{y}{x}=\sigma(\matind{A_1}{x}{y})/\sigma(1)$. Since
  $\sigma$ is a sesqui-morphism, the mapping $[x\mapsto
    \sigma(x)/\sigma(1)]$ is an automorphism and then
  $\rk(M') = \rk((A_1)^T) = rk(A_1)$. But, $\matgind{V_G\backslash
    X}{X} = \sigma(1) \cdot M'$. Then, $\rk(\matgind{V_G\backslash
    X}{X}) = \rk(M') = \rk(\matgind{X}{V_G\backslash X})$.

  For the second statement, we have by definition and Proposition
  \ref{prop:3.3},
    \begin{small}
    \begin{align*}
      \fcutrk_G(X) + \fcutrk_G(Y) &= \rk(A_1) + \rk(A_2)\\
      & \geq \rk(\matgind{X\cup Y}{V_G\backslash X \cap V_G\backslash Y}) +
      \rk(\matgind{X\cap Y}{V_G\backslash X\cup V_G\backslash
        Y}).
    \end{align*}
    \end{small}
    Since $V_G\backslash X\cap V_G\backslash Y = V_G\backslash (X\cup
    Y)$ and $V_G\backslash X \cup V_G\backslash Y= V_G\backslash
    (X\cap Y)$, the second statement holds. \qed
\end{pf*}


\begin{defn}[$\bF$-rank-width]\label{defn:3.2}  The
  \emph{$\bF$-rank-width} of a $\sigma$-symmetric
  $\bF^*$-graph $G$, denoted by $\frwd{G}$, is the
  $\fcutrk_G$-width of $V_G$.
\end{defn}

This definition generalises the one for undirected graphs. If we let
$\sigma_1$ be the identity automorphism on $\field{2}$, every
undirected graph is a $\sigma_1$-symmetric
$\field{2}^*$-graph. Moreover, for every undirected graph $G$, the
functions $\ucutrk_G$ and $\cutrk{\field{2}}_G$ are equal. It is then clear
that the definition of rank-width given in Section \ref{subsec:2.2}
coincides with the one of $\field{2}$-rank-width.

One can easily verify that the $\bF$-rank-width of a
$\sigma$-symmetric $\bF^*$-graph is the maximum of the
$\bF$-rank-width of its maximum connected components. The following
proposition, which says that $\bF$-rank-width and clique-width are
equivalent when $\bF$ is finite, has an easy proof. 
We omit it because its proof is an easy
adaptation of the one comparing rank-width and clique-width of
undirected graphs \cite[Proposition 6.3]{OS06}.

\begin{prop}\label{prop:3.4} Let $G$ be a $\sigma$-symmetric
  $\bF^*$-graph. Then, $\frwd{G}\leq \cwd{G} \leq 2\cdot
  q^{\frwd{G}}-1$.
\end{prop}

It is also easy to show that the clique-width and the $\bF$-rank-width are equivalent if $\bF$ is infinite but $C$ is finite. In \cite{KR09,KR10b}, authors present a decomposition related to $\bF$-rank-width and different characterizations of graphs of $\bF$-rank-width $1$.

\subsection{Vertex-Minor and Pivot-Minor}\label{subsec:3.2}

Bouchet generalised in \cite{BOU87} the notion of local
complementation to all graphs (undirected or not). We recall that a
graph $G$ is a $\field{2}^*$-graph and then is represented by a
$(V_G,V_G)$-matrix $\matg$ over $\field{2}$ where $\matgind{x}{y}:=1$
if and only if $(x,y)\in E_G$. A \emph{local complementation at $x$}
of $G$ is the graph represented by the matrix $\matg'$ over
$\field{2}$ where $\matind{M_G'}{z}{y} = \matgind{z}{y} +
\matgind{z}{x}\cdot \matgind{x}{y}$. This definition coincides with
the one on undirected graphs when $G$ is undirected. We will extend it
to $\bF^*$-graphs.  We say that $\lambda$ in $\bF^*$ is
\emph{$\sigma$-compatible} if $\sigma(\lambda) = \lambda\cdot \sigma(1)^2$.


\def\vz{z}
\def\vt{t}

\begin{defn}[$\lambda$-local complementation]\label{defn:3.3} Let
  $\lambda$ in $\bF^*$. Let $G$ be a $\bF^*$-graph and $x$ a vertex of
  $G$. The \emph{$\lambda$-local complementation at $x$} of $G$ is the
  $\bF^*$-graph $\lc{G}{x}$ represented by the $(V_G,V_G)$-matrix
  $\mat{\lc{G}{x}}$ where:
  \begin{align*}
    \matind{\mat{\lc{G}{x}}}{\vz}{\vt} &:= \begin{cases} \matgind{\vz}{\vt} +
      \lambda\cdot\matgind{\vz}{x}\cdot \matgind{x}{\vt} &
      \textrm{if $x\not\in \{\vz,\vt\}$},\\ \matgind{\vz}{\vt} &
      \textrm{otherwise}. \end{cases}
  \end{align*}
\end{defn}

One can easily verify that for every $\bF^*$-graph $G$ and every
vertex $x$ of $G$, the adjacency matrix of $\lc{G}{x}$ is obtained by
modifying the sub-matrix induced by the neighbors of $x$. Then for
every vertex $y$ of $G,\ \matgind{x}{y} = \matind{\mat{\lc{G}{x}}}{x}{y}$. 

\begin{defn}[locally equivalent, vertex-minor]\label{defn:loceq}
  A $\bF^*$-graph $H$ is \emph{locally equivalent} to a $\bF^*$-graph
  $G$ if $H$ is obtained by applying a sequence of $\lambda$-local
  complementations to $G$ with $\lambda\in\bF^*$.
  We call $H$ a \emph{vertex-minor} of $G$ if
  $H=G'[X]$ for some $X\subseteq V_G$ and $G'$ is locally equivalent
  to $G$.  Moreover, $H$ is a \emph{proper} vertex-minor of $G$ if 
  $X\subsetneq V_G$.
\end{defn}

In this section, we are interested in $\sigma$-symmetric graphs, thus
we have to restrict ourselves to a subset of local complementations
which preserve the $\sigma$-symmetry.  We now prove that
$\lambda$-local complementation is well defined on $\sigma$-symmetric
graphs when $\lambda$ is $\sigma$-compatible.

\begin{lem}\label{lem:3.2.1} 
Let $G$ be a $\sigma$-symmetric $\bF^*$-graph and let $\lambda\in \bF^*$ be $\sigma$-compatible. Then every $\lambda$-local complementation of $G$ is also $\sigma$-symmetric.
\end{lem}
\begin{pf*}{Proof.}
  Let $H:=\lc{G}{x}$ for some $\sigma$-compatible $\lambda$. It is sufficient to
  prove that $\matind{\mat{H}}{\vt}{\vz} =
  \sigma(\matind{\mat{H}}{\vz}{\vt})$ for any $\vz,\vt\in
  V_G,~\vz\ne \vt$.
  \begin{align*}
    \matind{\mat{H}}{\vt}{\vz} & = \matgind{\vt}{\vz} + \lambda\cdot
    \matgind{\vt}{x}\cdot \matgind{x}{\vz}\\ & =
    \sigma(\matgind{\vz}{\vt}) +
    \lambda\cdot\sigma(\matgind{x}{\vt})\cdot\sigma(\matgind{\vz}{x})\\ &=
    \sigma(\matgind{\vz}{\vt}) + \lambda\cdot
    \sigma(1)\cdot \sigma(\matgind{\vz}{x}\cdot \matgind{x}{\vt})\\
    & = \sigma(\matgind{\vz}{\vt}) +
    \sigma(\lambda)\cdot \sigma^{-1}(1)\cdot \sigma(\matgind{\vz}{x}\cdot \matgind{x}{\vt})\\ & =
    \sigma(\matgind{\vz}{\vt}) + \sigma(\lambda \cdot \matgind{\vz}{x}\cdot
    \matgind{x}{\vt})\\ &= \sigma(\matgind{\vz}{\vt} + \lambda \cdot \matgind{\vz}{x}\cdot
    \matgind{x}{\vt}) \\ & = \sigma(\matind{\mat{H}}{\vz}{\vt}).
    \end{align*}
\end{pf*}

\begin{defn}[$\sigma$-locally-equivalent, $\sigma$-vertex-minor]\label{defn:sloceq}
  A $\bF^*$-graph $H$ is \emph{$\sigma$-locally-equivalent} to a
  $\sigma$-symmetric $\bF^*$-graph $G$ if $H$ is obtained by applying
  a sequence of $\lambda$-local-complementations to $G$ with
  $\sigma$-compatibles $\lambda$.  We call $H$ a
  \emph{$\sigma$-vertex-minor} of $G$ if $H=G'[X]$ for some
  $X\subseteq V_G$ and $G'$ is $\sigma$-locally-equivalent to $G$.
  Moreover, $H$ is a \emph{proper} $\sigma$-vertex-minor of $G$ if
  $X\subsetneq V_G$.
\end{defn}

Note that if no $\sigma$-compatible $\lambda\in \bF^*$ exists, $H$ is a $\sigma$-vertex-minor of $G$ if and only if $H$ is an induced subgraph of $G$.

\begin{rem}\label{rem:3.1} Lemma \ref{lem:3.2.1} shows that
  $\lambda$-local complementation is well-defined on
  $\sigma$-symmetric $\bF^*$-graphs for $\sigma$-compatible
  $\lambda$. Moreover, one can easily verify that when $\bF$ is the
  field $\field{2}$, this notion of $1$-local complementation
  coincides with the one defined by Bouchet in \cite{BOU87}.
\end{rem}


The following lemma proves that $\sigma$-local-complementation does not increase $\bF$-rank-width.

\begin{lem} \label{lem:3.3} Let $G$ be a $\sigma$-symmetric
  $\bF^*$-graph and $x$ a vertex of $G$. For every subset $X$ of
  $V_G$,
  \begin{align*}
    \fcutrk_{\lc{G}{x}}(X) &= \fcutrk_G(X).
  \end{align*}
\end{lem}

\begin{pf*}{Proof.} We can assume that $x\in X$ since $\fcutrk_G$ is a
  symmetric function (Lemma \ref{lem:3.1}). For each $y\in X$, the
  $\sigma$-local-complementation at $x$ results in adding a multiple of the row
  indexed by $x$ to the row indexed by $y$. Precisely, we obtain
  $\matind{\mat{\lc{G}{x}}}{y}{V_G\backslash X}$ by adding $\lambda\cdot
  \matgind{y}{x}\cdot \matgind{x}{V_G\backslash X}$ to
  $\matgind{y}{V_G\backslash X}$. This operation is repeated for all
  $y\in X$. In each case, the rank of the matrix does not
  change. Hence, $\fcutrk_{\lc{G}{x}}(X) = \fcutrk_G(X)$.\qed
\end{pf*}

Unfortunately, such a $\sigma$-compatible $\lambda$ does not always exist. 
For instance, if the field is $\field{3}$ and $\sigma$ is such that
$\sigma(x)=-x$ (see Section~\ref{subsec:3.4}), no $\sigma$-compatible 
$\lambda$ does exist. We present now an other $\bF^*$-graph 
transformation which is defined for every couple $(\bF,\sigma)$.

\begin{defn}[Pivot-complementation]\label{defn:pivot} 
  Let $G$ be a $\sigma$-symmetric $\bF^*$-graph,
  and $x$ and $y$ two vertices of $G$ such that $\ell_G((x,y))\ne 0$. The
  \emph{pivot-complementation at $xy$} of $G$ is the
  $\bF^*$-graph $\pc{G}{xy}$ represented by the $(V_G,V_G)$-matrix
  $\mat{\pc{G}{xy}}$ where $\matind{\mat{\pc{G}{xy}}}{\vz}{\vz} := 0$ for every 
  $\vz\in V_G$, and for every $\vz,\vt\in V_G\setminus \{x,y\}$ with $\vz\ne \vt$:
  \begin{align*}
    \matind{\mat{\pc{G}{xy}}}{\vz}{\vt} &:= {\matgind{\vz}{\vt} -
    \frac{\matgind{\vz}{x}\cdot\matgind{y}{\vt}}{\matgind{y}{x}} -
    \frac{\matgind{\vz}{y}\cdot\matgind{x}{\vt}}{\matgind{x}{y}}}
  \end{align*}
  \begin{align*}
    \matind{\mat{\pc{G}{xy}}}{x}{\vt} &:= \frac{\matgind{y}{\vt}}{\matgind{y}{x}}
    &\matind{\mat{\pc{G}{xy}}}{y}{\vt} &:= \frac{\sigma(1)\cdot\matgind{x}{\vt}}{\matgind{x}{y}}\\
    \matind{\mat{\pc{G}{xy}}}{\vz}{x} &:= \frac{\sigma(1)\cdot\matgind{\vz}{y}}{\matgind{x}{y}}
    &\matind{\mat{\pc{G}{xy}}}{\vz}{y} &:= \frac{\matgind{\vz}{x}}{\matgind{y}{x}}\\
    \matind{\mat{\pc{G}{xy}}}{x}{y} &:= -\frac{1}{\matgind{y}{x}} 
    &\matind{\mat{\pc{G}{xy}}}{y}{x} &:= -\frac{\sigma(1)^2}{\matgind{x}{y}}
  \end{align*}
  A $\bF^*$-graph $H$ is \emph{pivot-equivalent} to a $\bF^*$-graph
  $G$ if $H$ is obtained by applying a sequence of 
  pivot-complementations to $G$.  
  We call $H$ a \emph{pivot-minor} of $G$
  if $H=G'[X]$ for some $X\subseteq V_G$ and $G'$ \emph{pivot-equivalent} to $G$.
  Moreover, $H$ is a \emph{proper} pivot-minor of $G$ if $X\subsetneq V_G$.
\end{defn}


Note that $\pc{G}{xy}=\pc{G}{yx}$ if $\sigma(1)=1$.
In the case of undirected graphs ($\bF=\field{2}$), this definition coincides with the pivot-complementation of undirected graphs~\cite{OUM05}. The following lemma shows that this transformation is well defined.

\begin{lem}\label{lem:3.2.2} 
Let $G$ be a $\sigma$-symmetric $\bF^*$-graph and let $xy$ be an edge of $G$. Then $\pc{G}{xy}$ is also $\sigma$-symmetric.
\end{lem}
\begin{pf*}{Proof.}
  Let $\vz,\vt\in V$, with $\vz\ne \vt$.
  If $\{\vz,\vt\}\cap\{x,y\}=\emptyset$, then 
  \begin{align*}
    \matind{\mat{\pc{G}{xy}}}{\vt}{\vz} & =  \matgind{\vt}{\vz} -
    \frac{\matgind{\vt}{x}\cdot\matgind{y}{\vz}}{\matgind{y}{x}} -
    \frac{\matgind{\vt}{y}\cdot\matgind{x}{\vz}}{\matgind{x}{y}} \\
    &= \sigma(\matgind{\vz}{\vt}) -
    \frac{\sigma(\matgind{x}{\vt})\cdot\sigma(\matgind{\vz}{y})}{\sigma(\matgind{x}{y})} -
    \frac{\sigma(\matgind{y}{\vt})\cdot\sigma(\matgind{\vz}{x})}{\sigma(\matgind{y}{x})} \\
    &= \sigma(\matgind{\vz}{\vt}) -
    \sigma \left (\frac{\matgind{x}{\vt}\cdot\matgind{\vz}{y}}{\matgind{x}{y}}\right ) -
    \sigma \left (\frac{\matgind{y}{\vt}\cdot\matgind{\vz}{x}}{\matgind{y}{x}}\right )\\ 
    &= \sigma \left(\matgind{\vz}{\vt}) -
    \frac{\matgind{x}{\vt}\cdot\matgind{\vz}{y}}{\matgind{x}{y}} -
    \frac{\matgind{y}{\vt}\cdot\matgind{\vz}{x}}{\matgind{y}{x}} \right )\\ 
    &= \sigma \left (\matind{\mat{\pc{G}{xy}}}{\vz}{\vt} \right).
  \end{align*}
  If $\vt\neq y$, then:
  \begin{align*}
    \matind{\mat{\pc{G}{xy}}}{\vt}{x} 
    &= \frac{\sigma(1)\cdot \matgind{\vt}{y}}{\matgind{x}{y}}
    = \frac{\sigma(1)\cdot \sigma(\matgind{y}{\vt})}{\sigma(\matgind{y}{x})}\\
    &= \sigma\left(\frac{\matgind{y}{\vt}}{\matgind{y}{x}}\right)
    = \sigma\left(\matind{\mat{\pc{G}{xy}}}{x}{\vt}\right).
  \end{align*}
  Finally:
  \begin{align*}
    \matind{\mat{\pc{G}{xy}}}{y}{x} &= -\frac{\sigma(1)^2}{\matgind{x}{y}}
    =-\frac{\sigma(1)^2}{\sigma(\matgind{y}{x})}\\
    &=\sigma \left ( - \frac{1^2}{\matgind{y}{x}} \right )
    =\sigma \left ( \matind{\mat{\pc{G}{xy}}}{x}{y} \right ).  ~\qed
\end{align*}
\end{pf*}

Similarly to Lemma~\ref{lem:3.3}, the following lemma
proves that pivot complementation does not increase $\bF$-rank-width.

\begin{lem} \label{lem:3.3.2} 
  Let $G$ be a $\sigma$-symmetric
  $\bF^*$-graph and $xy$ an edge of $G$. For every subset $X$ of
  $V_G$:
  $$\fcutrk_{\pc{G}{xy}}(X) = \fcutrk_G(X).$$
\end{lem}

\begin{pf*}{Proof.}
  Let $Y:=V_G\setminus X$. We can assume w.l.o.g. that $x\in X$.
  If $y\in X$, then  (with $X':=X\setminus \{x,y\}$)
  \begin{align*}
    \rk\left(\matind{\mat{\pc{G}{xy}}}{X}{Y} \right ) &=
    \rk\begin{pmatrix}
    \frac{1}{\matgind{y}{x}}\cdot \matgind{y}{Y}\\    
    \frac{\sigma(1)}{\matgind{x}{y}}\cdot \matgind{x}{Y}\\    
    \matgind{X'}{Y} - 
    \frac{\matgind{X'}{x}\cdot\matgind{y}{Y}}{\matgind{y}{x}} -
    \frac{\matgind{X'}{y}\cdot\matgind{x}{Y}}{\matgind{x}{y}}
    \end{pmatrix}\\
    &=
    \rk\begin{pmatrix}
    \frac{1}{\matgind{y}{x}}\cdot \matgind{y}{Y}\\    
    \frac{\sigma(1)}{\matgind{x}{y}}\cdot \matgind{x}{Y}\\    
    \matgind{X'}{Y} - 
    \frac{\matgind{X'}{x}\cdot\matgind{y}{Y}}{\matgind{y}{x}}
    \end{pmatrix}\\
    &=
    \rk\begin{pmatrix}
    \frac{1}{\matgind{y}{x}}\cdot \matgind{y}{Y}\\    
    \frac{\sigma(1)}{\matgind{x}{y}}\cdot \matgind{x}{Y}\\    
    \matgind{X'}{Y}
    \end{pmatrix}
    =
    \rk\begin{pmatrix}
    \matgind{y}{Y}\\    
    \matgind{x}{Y}\\    
    \matgind{X'}{Y}
    \end{pmatrix}\\
    &=
    \rk\left ( \matgind{X}{Y} \right ).
  \end{align*}
  If $y\not\in X$, then (with $X':=X\setminus \{x\}$ and $Y':=Y\setminus \{y\}$)
  \begin{align*}
    \rk\left(\matind{\mat{\pc{G}{xy}}}{X}{Y} \right ) &=
    \rk\begin{pmatrix}
    -\frac{1}{\matgind{y}{x}} & \frac{\matgind{y}{Y'}}{\matgind{y}{x}}\\
    \frac{\matgind{X'}{x}}{\matgind{y}{x}}     &    
    \matgind{X'}{Y'} - 
    \frac{\matgind{X'}{x}\cdot\matgind{y}{Y'}}{\matgind{y}{x}} -
    \frac{\matgind{X'}{y}\cdot\matgind{x}{Y'}}{\matgind{x}{y}}
    \end{pmatrix}\\
    &=
    \rk\begin{pmatrix}
    -\frac{1}{\matgind{y}{x}} & \frac{\matgind{y}{Y'}}{\matgind{y}{x}}\\
    0     &    
    \matgind{X'}{Y'} - 
    \frac{\matgind{X'}{y}\cdot\matgind{x}{Y'}}{\matgind{x}{y}}
    \end{pmatrix}\\
    &=
    \rk\begin{pmatrix}
    -\frac{1}{\matgind{y}{x}} & 0 \\
    0     &    
    \matgind{X'}{Y'} - 
    \frac{\matgind{X'}{y}\cdot\matgind{x}{Y'}}{\matgind{x}{y}}
    \end{pmatrix}\\
    &=
    \rk\begin{pmatrix}
    \matgind{x}{y} & 0 \\
    0     &    
    \matgind{X'}{Y'} - 
    \frac{\matgind{X'}{y}\cdot\matgind{x}{Y'}}{\matgind{x}{y}}
    \end{pmatrix}\\
    &=
    \rk\begin{pmatrix}
    \matgind{x}{y} & 0 \\
    \matgind{X'}{y}     &    
    \matgind{X'}{Y'} - 
    \frac{\matgind{X'}{y}\cdot\matgind{x}{Y'}}{\matgind{x}{y}}
    \end{pmatrix}\\
    &=
    \rk\begin{pmatrix}
    \matgind{x}{y} & \matgind{x}{Y'} \\
    \matgind{X'}{y}     &    
    \matgind{X'}{Y'} 
    \end{pmatrix}\\
    &= \rk\left ( \matgind{X}{Y} \right ).\qed
  \end{align*}
\end{pf*}

\begin{prop}\label{prop:3.5} Let $G$ and $H$ be two $\sigma$-symmetric
  $\bF^*$-graphs. If $H$ is $\sigma$-locally-equivalent
  (resp. pivot-equivalent) to $G$, then the $\bF$-rank-width of $H$ is
  equal to the $\bF$-rank-width of $G$. If $H$ is a
  $\sigma$-vertex-minor (resp. pivot-minor) of $G$, then the
  $\bF$-rank-width of $H$ is at most the $\bF$-rank-width of $G$.
\end{prop}

\begin{pf*}{Proof.} The first statement is obvious by Lemma
  \ref{lem:3.3} and Lemma~\ref{lem:3.3.2}. Since taking sub-matrices does not increase the rank,
  it does not increase the $\bF$-rank-width. So, the second statement
  is true.\qed
\end{pf*}

Our goal now is to prove the following which is a generalization of
Theorem \ref{thm:2.1}.

\begin{thm}\label{thm:3.1} 
\begin{enumerate}
\item[(i)]
  For each positive integer $k\geq 1$, there
  is a set $\sC_k^{(\bF,\sigma)}$ of $\sigma$-symmetric
  $\bF^*$-graphs, each having at most $(6^{k+1}-1)/5$ vertices, such
  that a $\sigma$-symmetric $\bF^*$-graph $G$ has $\bF$-rank-width at
  most $k$ if and only if no $\sigma$-symmetric $\bF^*$-graph in
  $\sC_k^{(\bF,\sigma)}$ is isomorphic to a pivot-minor of
  $G$.
\item[(ii)]
  Suppose that a $\sigma$-compatible $\lambda\in \bF^*$ exists.
  Then for each positive integer $k\geq 1$, there
  is a set ${\sC'}_k^{(\bF,\sigma)}$ of $\sigma$-symmetric
  $\bF^*$-graphs, each having at most $(6^{k+1}-1)/5$ vertices, such
  that a $\sigma$-symmetric $\bF^*$-graph $G$ has $\bF$-rank-width at
  most $k$ if and only if no $\sigma$-symmetric $\bF^*$-graph in
  ${\sC'}_k^{(\bF,\sigma)}$ is isomorphic to a $\sigma$-vertex-minor of
  $G$.
\end{enumerate}
\end{thm} 

Note that $\sC_k^{(\bF,\sigma)}$ and ${\sC'}_k^{(\bF,\sigma)}$ are finite
if $\bF$ is finite.
For doing so we adapt the same techniques as in
\cite{GGRW03,OUM05}. We first prove some inequalities concerning
cut-rank functions. The following one is a counterpart of
\cite[Proposition 4.3]{OUM05}. All the notions of linear algebra are
borrowed from \cite{LIP91}.

\begin{prop}\label{prop:3.6} Let $G$ be a $\sigma$-symmetric
  $\bF^*$-graph, $\lambda$ a $\sigma$-compatible element in $\bF^*$
  and $x$ a vertex of $G$.  For every subset $X$ of $V_G\setminus
  \{x\}$,
  \begin{small}
    \begin{align*}
      \fcutrk_{\subg{(\lc{G}{x})}{x}}(X) & = \rk \begin{pmatrix} -1 &
        \matgind{x}{V_G\backslash(X\cup x)} \\ \matgind{X}{x} &
        \matgind{X}{V_G\backslash (X\cup x)} \end{pmatrix} - 1
  \end{align*}\end{small}
\end{prop}

\begin{pf*}{Proof.} 
  Let $X$ be a subset
  of $V_G\backslash \{x\}$ and let $Y:=V_G\backslash (X\cup \{x\})$. We
  let $J$ be the matrix $(\matgind{\vz}{x}\cdot
  \matgind{x}{\vt})_{\vz\in X,\vt\in Y}$. Then,
  \begin{small}
    \begin{align*}
      \fcutrk_{\subg{(\lc{G}{x})}{x}}(X) 
      & = \rk (\matind{\mat{\lc{G}{x}}}{X}{Y})\\ 
      & = \rk (\matind{\mat{G}}{X}{Y}+\lambda\cdot J)\\ 
      & = \rk
      \underbrace{\begin{pmatrix} -1\cdot \lambda^{-1} &&
          \matgind{x}{Y}  \\ 0 &&
          \matgind{X}{Y} + \lambda\cdot J 
      \end{pmatrix}}_{A} - 1
  \end{align*}\end{small}
  We now show how to transform the $( \{x\}\cup X, \{x\}\cup
  Y)$-matrix $A$ by using elementary row operations in order to get
  the desired equality.  For each $\vz\in X$, 
  \begin{small}\begin{align*} -\lambda\cdot \matgind{\vz}{x}\cdot \matind{A}{x}{Y\cup \{x\}} 
      &= \begin{pmatrix}
        \matgind{\vz}{x} && -\lambda\cdot \matind{J}{\vz}{Y} 
      \end{pmatrix}.\end{align*}\end{small} Hence, 
  \begin{small} \begin{align*}-\lambda\cdot \matgind{\vz}{x}\cdot \matind{A}{x}{Y\cup \{x\}} +
      \matind{A}{\vz}{Y\cup \{x\}} = \begin{pmatrix} \matgind{\vz}{x}&&
        \matgind{\vz}{Y}\end{pmatrix}.\end{align*}\end{small}
  Therefore, by adding $-\lambda\cdot \matgind{\vz}{x}\cdot \matind{A}{x}{Y\cup
    \{x\}}$ to each row $\matind{A}{\vz}{Y\cup \{x\}}$ of $A$ we get
  the matrix {\small $\begin{pmatrix} -1 & \matgind{x}{Y}
      \\ \matgind{X}{x} & \matgind{X}{Y} \end{pmatrix}$}. This
  concludes the proof. \qed
\end{pf*}

The following lemma is thus the counterpart of \cite[Lemma
  4.4]{OUM05} and \cite[Proposition 3.2]{GGRW03}. 

\begin{lem}\label{lem:3.4} Let $G$ be a
  $\sigma$-symmetric $\bF^*$-graph and $x$ a vertex in $V_G$.  Assume
  that $(X_1,X_2)$ and $(Y_1,Y_2)$ are partitions of
  $V_G\backslash \{x\}$. Then, \begin{small} \begin{align*} \fcutrk_{\subg{G}{x}}(X_1) +
      \fcutrk_{\subg{(\lc{G}{x})}{x}}(Y_1) \geq \fcutrk_G(X_1\cap Y_1) +
      \fcutrk_G(X_2\cap Y_2) - 1. \end{align*}\end{small}
\end{lem}

\begin{pf*}{Proof.} We recall that for every vertex $z$ of $G$,
  $\matgind{z}{z}=0$. Let $M'$ be obtained from $\matg$ by replacing
  $\matgind{x}{x}$ by $-1$. It is worth noticing that for every subset
  $X$ of $V_G$, $\rk(\matgind{X}{V_G\backslash X}) =
  \rk(\matind{M'}{X}{V_G\backslash X}$. We recall that
  $Y_2=V_G\backslash(Y_1\cup \{x\})$ and $X_2=V_G\backslash (X_1\cup
  \{x\})$. By definition of $M'$, \begin{small} \begin{align*} \matind{M'}{Y_1\cup \{x\}}{Y_2\cup
        \{x\}}&=\begin{pmatrix} -1 & \matgind{x}{Y_2} \\ \matgind{Y_1}{x} &
      \matgind{Y_1}{Y_2} \end{pmatrix}.\end{align*}\end{small}
  By Proposition \ref{prop:3.6}, 
  \begin{align*} \fcutrk_{\subg{G}{x}}(X_1) +
    \fcutrk_{\subg{(\lc{G}{x})}{x}}(Y_1) = \rk (\matgind{X_1}{X_2}) + \rk
    (\matind{M'}{Y_1\cup 
      \{x\}}{Y_2\cup \{x\}}) - 1.
  \end{align*}
  Since $\rk (\matgind{X_1}{X_2}) = \rk(\matind{M'}{X_1}{X_2})$, by
  Proposition \ref{prop:3.3} we get the inequality {\small
    \begin{align*} \rk (\matgind{X_1}{X_2}) + \rk (\matind{M'}{Y_1\cup
        \{x\}}{Y_2\cup \{x\}}) \geq\\ \rk(\matind{M'}{X_1\cap
        Y_1}{X_2\cup Y_2\cup \{x\}}) + \rk (\matind{M'}{X_1\cup Y_1 \cup
        \{x\}}{X_2\cap Y_2}).\end{align*}} Hence, {\small \begin{align*}
      \fcutrk_{\subg{G}{x}}(X_1) + \fcutrk_{\subg{(\lc{G}{x})}{x}}(Y_1) \geq
      \fcutrk_{G}(X_1\cap Y_1) + \fcutrk_{G}(X_1\cup Y_1\cup x) -
      1.\end{align*}} By the symmetry of $\fcutrk_G$, we get the desired
  inequality.\qed
\end{pf*}

Similarly, we get the followings for pivot-minor.

\begin{prop}\label{prop:3.6.2} Let $G=(V,E)$ be a $\sigma$-symmetric
  $\bF^*$-graph and $xy$ an edge of $G$. For every subset $X$ of 
  $V_G\setminus \{x\}$,
  \begin{small}
    \begin{align*}
      \fcutrk_{\subg{(\pc{G}{xy})}{x}}(X) & = \rk 
      \begin{pmatrix} 0 & \matgind{x}{V\backslash(X\cup x)} \\ 
        \matgind{X}{x} & \matgind{X}{V\backslash (X\cup x)} 
      \end{pmatrix} - 1
  \end{align*}\end{small}
\end{prop}

\begin{pf*}{Proof.} 
  Suppose w.l.o.g. that $y\in X$ (otherwise replace $X$ by $V_G\setminus (X\cup \{x\})$).
  Let $Y:=V_G\setminus (X\cup \{x\})$ and $X':=X\setminus \{y\}$.
  Then, by elementary row and column operations, we have: 
  \begin{align*}
    \fcutrk_{\subg{(\pc{G}{xy})}{x}}(X) & = 
    \rk\begin{pmatrix}
    \frac{\sigma(1)}{\matgind{x}{y}}\cdot\matgind{x}{Y}\\
    \matgind{X'}{Y} - 
    \frac{\matgind{X'}{x}\cdot\matgind{y}{Y}}{\matgind{y}{x}} -
    \frac{\matgind{X'}{y}\cdot\matgind{x}{Y}}{\matgind{x}{y}}
    \end{pmatrix}\\
    &=
    \rk\begin{pmatrix}
    \matgind{y}{Y}\\
    \matgind{X'}{Y} - 
    \frac{\matgind{X'}{x}\cdot\matgind{y}{Y}}{\matgind{y}{x}}
    \end{pmatrix}\\
    &=
    \rk\begin{pmatrix}
    \matgind{y}{x} & \matgind{y}{Y} \\
    0 &\matgind{y}{Y}\\
    0 &\matgind{X'}{Y} - 
    \frac{\matgind{X'}{x}\cdot\matgind{y}{Y}}{\matgind{y}{x}}
    \end{pmatrix}-1\\
    &=
    \rk\begin{pmatrix}
    0 &\matgind{y}{Y}\\
    \matgind{y}{x} & \matgind{y}{Y} \\
    \matgind{X'}{x} &\matgind{X'}{Y}
    \end{pmatrix}-1\\
    &=
    \rk\begin{pmatrix}
    0 &\matgind{y}{Y}\\
    \matgind{X}{x} &\matgind{X}{Y} 
    \end{pmatrix}-1. \qed 
  \end{align*}
\end{pf*}

\begin{lem}\label{lem:3.4.2} Let $G$ be a
  $\sigma$-symmetric $\bF^*$-graph and $xy$ an edge in $V_G$.  Assume
  that $(X_1,X_2)$ and $(Y_1,Y_2)$ are partitions of
  $V_G\backslash \{x\}$. Then 
  \begin{small} 
    \begin{align*} 
      \fcutrk_{\subg{G}{x}}(X_1) +
      \fcutrk_{\subg{(\pc{G}{xy})}{x}}(Y_1) \geq \fcutrk_G(X_1\cap Y_1) +
      \fcutrk_G(X_2\cap Y_2) - 1. 
    \end{align*}
  \end{small}
\end{lem}

\begin{pf*}{Proof.} 
  We recall that
  $Y_2=V_G\backslash(Y_1\cup \{x\})$ and $X_2=V_G\backslash (X_1\cup
  \{x\})$. By definition of $M$, 
  \begin{small} 
    \begin{align*} 
      \matind{M}{Y_1\cup \{x\}}{Y_2\cup \{x\}}
      &=\begin{pmatrix}0 & \matgind{x}{Y_2} \\ 
      \matgind{Y_1}{x} & \matgind{Y_1}{Y_2} \end{pmatrix}.
    \end{align*}
  \end{small}
  By Proposition \ref{prop:3.6.2}, 
  \begin{align*} \fcutrk_{\subg{G}{x}}(X_1) +
    \fcutrk_{\subg{(\pc{G}{x})}{xy}}(Y_1) = \rk (\matgind{X_1}{X_2}) + \rk
    (\matind{M}{Y_1\cup 
      \{x\}}{Y_2\cup \{x\}}) - 1.
  \end{align*}
  By Proposition \ref{prop:3.3} we get the inequality 
  {\small
    \begin{align*} \rk (\matgind{X_1}{X_2}) + \rk (\matind{M}{Y_1\cup
        \{x\}}{Y_2\cup \{x\}}) \geq\\ 
      \rk(\matind{M}{X_1\cap
        Y_1}{X_2\cup Y_2\cup \{x\}}) + \rk (\matind{M}{X_1\cup Y_1 \cup
        \{x\}}{X_2\cap Y_2}).
  \end{align*}
  } 
  Hence, 
  {\small 
    \begin{align*}
      \fcutrk_{\subg{G}{x}}(X_1) + \fcutrk_{\subg{(\pc{G}{xy})}{x}}(Y_1) \geq
      \fcutrk_{G}(X_1\cap Y_1) + \fcutrk_{G}(X_1\cup Y_1\cup x) -
      1.
  \end{align*}
  } 
  By the symmetry of $\fcutrk_G$, we get the desired inequality.\qed
\end{pf*}

The most important ingredients for proving Theorem \ref{thm:3.1} are
Propositions \ref{prop:3.6} and \ref{prop:3.6.2}, and Lemmas
\ref{lem:3.4} and \ref{lem:3.4.2}. All the other ingredients are
already proved in \cite{GGRW03,OUM05} except that they are stated for
the connectivity function of matroids in \cite{GGRW03} and for
undirected graphs in \cite{OUM05}. Their proofs rely only on the fact
that the parameter is symmetric, submodular and integer valued. We
include them for completeness. We first recall some definitions
\cite{GGRW03,OUM05}.

Let $V$ be a finite set and $f:2^V\to \bN$ a symmetric and submodular
function. Let $(A,B)$ be a bipartition of $V$. A \emph{branching of}
$B$ is a triple $(T,r,\cL)$ where $T$ is a sub-cubic tree with a fixed
node $r\in \leaves{T}$ and such that $(\subg{T}{r},\cL)$ is a layout
of $B$. For an edge $e$ of $T$ and a node $v$ of $T$, we let $T_{ev}$
be the set of nodes in the component of $\subg{T}{e}$ not containing
$v$ and we let $Y_{ev}:=\cL^{-1}(\leaves{T_{ev}})$. We say that $B$ is
\emph{$k$-branched} if there exists a branching $(T,r,\cL)$ such that
for each edge $e$ of $T$, $f(Y_{er}) \leq k$. It is worth noticing
that if $A$ and $B$ are $k$-branched, then the $f$-width of $V$ is at
most $k$.

A subset $A$ of $V$ is called \emph{titanic} with respect to $f$ if for every partition $(A_1,A_2,A_3)$ of $A$, there is a $i\in \{1,2,3\}$ such that $f(A_i) \ge f(A)$ ($A_1$,$A_2$ or $A_3$ may be empty). 

The following lemma is proved in \cite[Lemma 5.1]{OUM05} for
$\ucutrk_G$, in \cite[Lemma 2.1]{GGRW03} for the connectivity function
of matroids, and in \cite[Lemma 3.3]{HO07} for all symmetric and
submodular functions.

\begin{lem}[{\cite[Lemma 3.3]{HO07}}]\label{lem:3.5}
  Let $V$ be a finite set and $f:2^V\to \bN$ a symmetric and
  submodular function. Assume that the $f$-width of $V$ is at most $k$.
  Let $(A,B)$ be a bipartition of $V$ such that $f(A) \leq k$. If $A$
  is titanic with respect to $f$, then $B$ is $k$-branched.
\end{lem}

Let $g:\bN\to \bN$ be a function. A $\sigma$-symmetric $\bF^*$-graph
$G$ is called \emph{$(m,g)$-connected} if for every bipartition
$(A,B)$ of $V_G$, $\fcutrk_G(A) = \ell < m$ implies $|A|\leq
g(\ell)$ or $|B|\leq g(\ell)$. This notion will help to bound the
order of the minimal $\sigma$-symmetric $\bF^*$-graphs that every
$\sigma$-symmetric $\bF^*$-graph of $\bF$-rank-width $k$ must
exclude as pivot-minor or $\sigma$-vertex-minors.

\begin{lem}\label{lem:3.6} Let $f:\bN\to \bN$ be a
  non-decreasing function with $f(0)=0$. Let $G$ be an $(m,f)$-connected
  $\sigma$-symmetric $\bF^*$-graph and $x$ a vertex of $G$. 
  Then either $\subg{G}{x}$  or  $\subg{(\pc{G}{xy})}{x}$ is $(m,2f)$-connected (for an edge $xy$).
  Moreover if a $\sigma$-compatible $\lambda\in\bF^*$ exists, either $\subg{G}{x}$ or $\subg{(\lc{G}{x})}{x}$ is $(m,2f)$-connected.
\end{lem}

\begin{pf*}{Proof.} 
  Since $f(0)=0$, $G$ is connected. Let $y$ be a neighbor of $x$.
  Suppose neither $\subg{G}{x}$ nor
  $\subg{(\pc{G}{xy})}{x}$ is $(m,2f)$-connected. Then there are
  bipartitions $(A_1,A_2)$ and $(B_1,B_2)$ of $V_G\backslash \{x\}$
  such that $a = \fcutrk_{\subg{G}{x}}(A_1)$, $b =
  \fcutrk_{\subg{(\pc{G}{xy})}{x}}(B_1)$, and  $|A_i| > 2f(a)$, $|B_i| >
  2f(b)$ for $i=1,2$.
  
  We may assume that $a\geq b$ 
  . By
  Lemma \ref{lem:3.4.2}, we have
  \begin{align*}
    \fcutrk_G(A_1\cap B_1) +\fcutrk_G(A_2\cap B_2) \leq a +b +1.
  \end{align*}
  Thus, either $\fcutrk_G(A_1\cap B_1) \leq a$ or $\fcutrk_G(A_2\cap
  B_2) \leq b$. So, by hypothesis either $|A_1\cap B_1|\leq f(a)$ or
  $|A_2\cap B_2|\leq f(b)$. Assume that $|A_2\cap B_2| \leq
  f(b)$. Similarly, we also have either $|A_2\cap B_1|\leq f(a)$ or
  $|A_1\cap B_2|\leq f(b)$. Since $|A_1\cap B_2| = |B_2|-|B_2\cap A_2|
  > f(b)$, we have $|A_2\cap B_1|\leq f(a)$. Then $|A_2| = |A_2\cap
  B_1| + |A_2\cap B_2| \leq f(a) + f(b) \leq 2 f(a)$. This yields a
  contradiction.
  
  The proof of the second statement is similar, using Lemma \ref{lem:3.4}.
  \qed
\end{pf*}

  We let $g(n) = (6^n-1)/5$. Note that $g(0) = 0, ~g(1) = 1$ and $g(n) =
  6g(n-1)+1$ for all $n\geq 1$. We now prove that the minimal
  $\sigma$-symmetric $\bF^*$-graphs that have $\bF$-rank-width at
  least $k+1$ are $(k+1,g)$-connected.

  \begin{lem} \label{lem:3.7}  Let $k\geq 1$ and let $G$ be a
    $\sigma$-symmetric $\bF^*$-graph of $\bF$-rank-width larger than $k$. 
    If every proper pivot-minor of $G$
    has $\bF$-rank-width at most $k$, then $G$ is $(k+1,g)$-connected.
    Similarly, if a $\sigma$-compatible $\lambda\in \bF^*$ exists,  
    and  every proper $\sigma$-vertex-minor of $G$
    has $\bF$-rank-width at most $k$, then $G$ is $(k+1,g)$-connected.
  \end{lem}

  \begin{pf*}{Proof.} The proof is similar to the one of \cite[Lemma
      5.3]{OUM05}. We assume that $G$ is connected since the $\bF$-rank-width of
    $G$ is the maximum of the $\bF$-rank-width of its connected
    components. It is now easy to see that $G$ is $(1,g)$-connected.

    Assume that $m\leq k$ and that $G$ is $(m,g)$-connected but $G$ is
    not $(m+1,g)$-connected. Then there exists a bipartition $(A,B)$
    with $\fcutrk_G(A) = m$ such that $|A|>g(m)$ and $|B|>g(m)$. Also,
    either $A$ or $B$ is not $k$-branched ($\frwd{G} > k$). We may
    assume that $B$ is not $k$-branched. Let $x\in A$ and $xy\in E_G$.

    By Lemma \ref{lem:3.6}, either $\subg{G}{x}$ or $\subg{(\pc{G}{xy})}{x}$
    is $(m,2g)$-connected; assume $\subg{G}{x}$ is
    $(m,2g)$-connected. Since $\subg{G}{x}$ and $\subg{(\pc{G}{xy})}{x}$ are
    proper pivot-minors of $G$, they both have $\bF$-rank-width at
    most $k$. Let $(A_1,A_2,A_3)$ be a tri-partition of $A\backslash
    \{x\}$. Since $|A|>g(m) = 6g(m-1)+1$, there exists an $i\in [3]$
    such that $|A_i|>2g(m-1)$. Since $\subg{G}{x}$ is
    $(m,2g)$-connected and $|A_i| > 2g(m-1)$, \begin{align*}
      \fcutrk_{\subg{G}{x}}(A_i) \geq m \geq
      \fcutrk_{\subg{G}{x}}(A\backslash \{x\}). \end{align*}
    Therefore, by Lemma \ref{lem:3.5} $B$ is $k$-branched in
    $\subg{G}{x}$. Since $B$ is not $k$-branched in $G$, there exists
    $W\subseteq B$ such that \begin{align*} \fcutrk_G(W) =
      \fcutrk_{\subg{G}{x}}(W) + 1.\end{align*}
    Thus, the column vectors of $\matgind{W}{V_G\backslash (W\cup
      \{x\})}$ do not span $\matgind{W}{x}$. So, the column vectors of
    $\matgind{W}{V_G\backslash (B\cup \{x\})}$ do not span
    $\matgind{W}{x}$. Hence, the column vectors of
    $\matgind{B}{V_G\backslash (B\cup \{x\})}$ do not span
    $\matgind{B}{x}$. Therefore, \begin{align*} \fcutrk_{\subg{G}{x}}(B) =
      \fcutrk_G(B) - 1 = m - 1. \end{align*} This implies that $|B| \leq
    2g(m-1)$ or $|A\backslash\{x\}| \leq 2g(m-1)$. A contradiction.

    The proof of the second statement is similar (replace $\pc{G}{xy}$ by $\lc{G}{x}$).\qed
  \end{pf*}

  As a consequence of Lemma \ref{lem:3.7}, we get the following.

  \begin{thm}[Size of Excluded Pivot-Minor and $\sigma$-Vertex-Minors]\label{thm:3.2} 
    Let $k\geq
    1$ and let $G$ be a $\sigma$-symmetric $\bF^*$-graph. If $G$ has
    $\bF$-rank-width larger than $k$ but every proper pivot-minor of
    $G$ has $\bF$-rank-width at most $k$, then $|V_G|\leq
    (6^{k+1}-1)/5$.
    
    Moreover, if a $\sigma$-compatible $\lambda\in\bF^*$ exists, and if $G$ has
    $\bF$-rank-width larger than $k$ but every proper $\sigma$-vertex-minor of
    $G$ has $\bF$-rank-width at most $k$, then $|V_G|\leq
    (6^{k+1}-1)/5$.
  \end{thm}
  
  \begin{pf*}{Proof.} Let $x\in V_G$. We may assume that $\subg{G}{x}$ is
    $(k+1,2g)$-connected by Lemmas \ref{lem:3.6} and
    \ref{lem:3.7}. Since $\subg{G}{x}$ has $\bF$-rank-width $k$, there
    exists a bipartition $(A,B)$ of $V_G\backslash\{x\}$ such that $|A|\geq
    \frac{1}{3}(|V_G|-1)$ and $|B|\geq \frac{1}{3}(|V_G|-1)$ and
    $\fcutrk_{\subg{G}{x}}(A) \leq k$. By $(k+1,2g)$-connectivity, either
    $|A|\leq 2g(k)$ or $|B|\leq 2g(k)$. Therefore, $|V_G|-1 \leq 6g(k)$ and
    consequently $|V_G|\leq 6g(k)+1 = g(k+1)$.\qed
  \end{pf*}

  It is surprising that the bound $(6^{k+1}-1)/5$ does not depend
  neither on $\bF$ nor on $\sigma$. But that is because the proof
  technique is based on the $\fcutrk_G$-width of $V_G$ and neither on
  $\bF$ nor on $\sigma$. However, the $\bF$-rank-width depends on
  $\bF$ since there is no reason that the rank of a matrix is the same
  in two different fields. But, as we will see in the following proof
  of Theorem \ref{thm:3.1}, the set of $\sigma$-symmetric
  $\bF^*$-graphs to exclude as pivot-minors and $\sigma$-vertex-minor 
  depends on $\bF$ and $\sigma$.

  \begin{pf*}{Proof of Theorem \ref{thm:3.1}.}  
    We show only the proof for the first statement. the other proof is similar.
    If $k< 0$, we let $\sC_k^{(\bF,\sigma)}
    = \emptyset$. If $ k = 0$, we let $\sC_0^{(\bF,\sigma)} :=
    \{\const{a}~|~a\in \bF^*\}$ where $\const{a}$ is the
    $\sigma$-symmetric $\bF^*$-graph $(\{x,y\}, \{x\overset{a}\to
    y,\ y\overset{\sigma(a)}\to x\})$. It is clear that $G$ has
    $\bF$-rank-width at most $0$ if and only if $G$ has no
    pivot-minor isomorphic to any $\const{a} \in
    \sC_0^{(\bF,\sigma)}$.

    Assume now that $k\geq 1$ and let $\sC_k^{(\bF,\sigma)}$ be the
    set, up to isomorphism, of $\sigma$-symmetric
    $\bF^*$-graphs $H$ such that $\frwd{H} >k$ and every proper
    pivot-minor of $H$ has $\bF$-rank-width at most $k$. By Theorem
    \ref{thm:3.2}, each
    $\sigma$-symmetric $\bF^*$-graph in $\sC_k^{(\bF,\sigma)}$ has at
    most $(6^{k+1}-1)/5$ vertices.

    Let $G$ be a $\sigma$-symmetric $\bF^*$-graph of $\bF$-rank-width
    at most $k$. Since every $\bF^*$-graph in $\sC_k^{(\bF,\sigma)}$
    has $\bF$-rank-width larger than $k$, no $\bF^*$-graph in
    $\sC_k^{(\bF,\sigma)}$ is isomorphic to a pivot-minor of $G$.

    Conversely,  assume that the $\bF$-rank-width of $G$ is larger than $k$ and
    let $H$ be a proper pivot-minor of $G$ of minimum size such that
    $\frwd{H} > k$. Then there exists a $\bF^*$-graph $H'\in
    \sC_k^{(\bF,\sigma)}$ isomorphic to $H$.\qed
  \end{pf*}



  
Moreover, using the characterization of $\bF^*$-graphs of $\bF$-rank-width $1$ \cite{KR09,KR10b}, obstructions for $\bF^*$-graphs of $\bF$-rank-width $1$ by vertex-minor (resp. pivot-minor) have at most $5$ (resp. $6$) vertices.
In \cite{OUM05b}, Oum derives from the \emph{principal pivot
  transformation} of Tucker (see \cite{TSAT00} for instance) a notion
of \emph{pivot-minor} for symmetric and skew-symmetric matrices and
proved that symmetric and skew-symmetric matrices of bounded
rank-width are well-quasi-ordered by this relation. Our notion of
pivot-minor is a special case of Oum's notion when $\sigma(x):=x$ or $\sigma(x):=-x$. Hence, oriented graphs
of bounded rank-width are well-quasi-ordered by pivot-minor. We
generalise Oum's result to $\sigma$-symmetric matrices in \cite{KR10}.



\subsection{Recognizing $\bF$-Rank-Width at Most $k$}\label{subsec:3.3}

We give in this section a cubic-time algorithm that decides whether a
$\bF^*$-graph has $\bF$-rank-width at most $k$, for fixed finite field 
$\bF$ and a fixed $k$. This
algorithm is an easy corollary of the one by \hlineny and Oum
concerning representable matroids \cite{HO07}. We recall the necessary
materials about matroids. We refer to Schrijver \cite{SCH03} for our
matroid terminology. We let $\bF$ be a fixed finite field and
$\sigma:\bF\to \bF$ a sesqui-morphism. 

\begin{defn}[Matroids]\label{defn:3.4}  A pair $\cM=(S,\cI)$ is called a
  \emph{matroid} if $S$ is a finite set and $\cI$ is a nonempty
  collection of subsets of $S$ satisfying the following conditions
  \begin{enumerate}
  \item[(M1)] if $I\in \cI$ and $J\subseteq I$, then $J\in \cI$,
  \item[(M2)] if $I,J\in \cI$ and $|I|<|J|$, then $I\cup \{z\}\in \cI$ for
    some $z\in J\setminus I$.
  \end{enumerate}
  
  For $U\subseteq S$, a subset $B$ of $U$ is called a \emph{base} of
  $U$ if $B$ is an inclusionwise maximal subset of $U$ and belongs to
  $\cI$. It is easy to see that, if $B_1$ and $B_2$ are bases of
  $U\subseteq S$, then $B_1$ and $B_2$ have the same size. The common
  size of the bases of a subset $U$ of $S$ is called the \emph{rank of
    $U$}, denoted by $r_{\cM}(U)$. A set $B\subseteq S$ is a base of $\cM$
  if it is a base of $S$.
  
  Let $A$ be a $m\times n$-matrix. Let $S:=\{1,\ldots, n\}$ and let
  $\cI$ be the collection of all those subsets $I$ of $S$ such that
  the columns of $A$ with index in $\cI$ are linearly independent. Then
  $\cM:=(S,\cI)$ is a matroid. If $A$ has entries in $\bF$, then $\cM$ is
  said \emph{representable over $\bF$} and $A$ is called a
  \emph{representation of $\cM$ over $\bF$}.
  
  We now define the branch-width of matroids. Let $\cM=(S,\cI)$ be a
  matroid. We let $\lambda_{\cM}$ be defined such that for every subset
  $U$ of $S,\ \lambda_{\cM}(U) = r_{\cM}(U) + r_{\cM}(S\backslash U) -r_{\cM}(S)+ 1$
  and call it the \emph{connectivity function of $\cM$}. The function
  $\lambda_{\cM}$ is symmetric and submodular \cite{SCH03}. The
  \emph{branch-width of $\cM$}, denoted by $\bwd{\cM}$, is the
  $\lambda_{\cM}$-width of $S$.
\end{defn}

\begin{defn}[Partitioned Matroids \cite{HO07}]\label{defn:3.5} Let
  $\cM=(S,\cI)$ be a matroid and $\cP$ a partition of $S$. The couple
  $(\cM,\cP)$ is called a \emph{partitioned matroid}. A partitioned
  matroid $(\cM,\cP)$ is \emph{representable over $\bF$} if $\cM$ is
  representable over $\bF$. For a partitioned matroid $(\cM,\cP)$, we
  let $\lambda_{\cM}^{\cP}$ be defined such that for every $Z\subseteq
  \cP$, we have $\lambda_{\cM}^{\cP}(Z) := \lambda_{\cM}(\bigcup_{Y\in Z}
  Y)$. The \emph{branch-width of $(\cM,\cP)$}, denoted by $\bwd{\cM,\cP}$,
  is the $\lambda_{\cM}^{\cP}$-width of $\cP$.
\end{defn}

We recall the following important result by \hlineny and Oum
\cite{HO07}.

\begin{thm}[\cite{HO07}] \label{thm:3.3} Let $\bF$ be a fixed finite field, 
  and $k$ be a fixed positive
  integer. There exists a cubic-time algorithm that takes as input a
  representable partitioned matroid $(\cM,\cP)$ over $\bF$ given with
  the representation of $\cM$ over $\bF$ and outputs a layout of $\cP$
  of $\lambda_{\cM}^{\cP}$-width at most $k$ or confirms that the
  branch-width of $(\cM,\cP)$ is strictly greater than $k$.
\end{thm}

We can now derive our recognition algorithm from Theorem
\ref{thm:3.3}. For that we borrow ideas from \cite{HO07}. For a set
$X$, we let $X'$ be a disjoint copy of it defined as $\{x'\mid x\in
X\}$. For $G$ a $\bF^*$-graph, we let $\cM_G$ be the matroid on
$V_G\cup V_G'$ represented by the $(V_G,V_G\cup V_G')$-matrix (recall
that $I_n$ denotes the identity square matrix of size $n$):

\begin{align*}
  \begin{matrix}
    & V_G\qquad V_G'\\
    V_G & \Big(\begin{matrix}  I_{|V_G|} \qquad  \matg\end{matrix}\Big)
  \end{matrix}
\end{align*}

For each $x\in V$, we let $P_x:=\{x,x'\}$ and we let
$\Pi(G):=\{P_x~|~x\in V_G\}$. We now prove the following which is a
counterpart of \cite[Proposition 3.1]{OUM05}.

\begin{prop}\label{prop:3.7} Let $G$ be a $\bF^*$-graph. For every
  $X\subseteq V_G,\ \lambda_{\cM_G}^{\Pi(G)}(P) =
  \rk(\matgind{X}{V_G\backslash X}) + \rk(\matgind{V_G\backslash X}{X})
  +1$ where $P:=\{P_x~|~x\in X\}$.
\end{prop}

\begin{pf*}{Proof.} For $X\subseteq V_G$ and $P:=\{P_x~|~x\in X\}$, we
  have
  \begin{align*}
    \lambda_{\cM_G}^{\Pi(G)}(P) & = r_{\cM_G}(X\cup X') +
    r_{\cM_G}(V_G\backslash X\cup
    (V_G\backslash X)') - r_{\cM_G}(V_G\cup V_G') + 1 \\ & = rk \begin{pmatrix} 0 &
      \matgind{V_G\backslash X}{X} \\ I_{|X|} & \matgind{X}{X} \end{pmatrix} +
    rk \begin{pmatrix} 0 & \matgind{X}{V_G\backslash X} \\ I_{|V_G|-|X|} &
      \matgind{V_G\backslash X}{V_G\backslash X} \end{pmatrix} -|V_G| + 1\\ & = |X| +
    \rk(\matgind{V_G\backslash X}{X}) + |V_G-X| +
    \rk(\matgind{X}{V_G\backslash X}) -|V_G| + 1\\ & =
    \rk(\matgind{X}{V_G\backslash X}) + \rk(\matgind{V_G\backslash X}{X})
    +1. \qed \end{align*}
\end{pf*}

Since when $G$ is $\sigma$-symmetric, we have
$\rk(\matgind{X}{V_G\backslash X}) = \rk(\matgind{V_G\backslash X}{X})
= \fcutrk_G(X)$, we get the followings as corollaries of Proposition
\ref{prop:3.7}.

\begin{cor}\label{cor:3.1} Let $G$ be a $\sigma$-symmetric
  $\bF^*$-graph. For every $X\subseteq
  V_G,\ \lambda_{\cM_G}^{\Pi(G)}(P) =2\cdot \fcutrk_G(X) +1$ where
  $P:=\{P_x~|~x\in X\}$.
\end{cor}

\begin{cor}\label{cor:3.2} Let $G$ be a $\sigma$-symmetric
  $\bF^*$-graph and let $p:V_G\to \Pi(G)$ be the bijective function
  such that $p(x)=P_x$. If $(T,\cL)$ is a layout of $\Pi(G)$ of
  $\lambda_{\cM_G}^{\Pi(G)}$-width $2k+1$, then $(T,\cL\circ p)$ is a
  layout of $V_G$ of $\fcutrk_G$-width $k$. Conversely, if $(T,\cL)$
  is a layout of $V_G$ of $\fcutrk_G$-width $k$, then $(T,\cL\circ
  p^{-1})$ is a layout of $\Pi(G)$ of $\lambda_{\cM_G}^{\Pi(G)}$-width
  $2k+1$.
\end{cor}

\begin{thm}[Checking $\bF$-Rank-Width at most $k$]\label{thm:3.4}
  For fixed $k$ and a fixed finite field $\bF$, there exists a 
  cubic-time algorithm that, for a
  $\sigma$-symmetric $\bF^*$-graph $G$, either outputs a layout of
  $V_G$ of $\fcutrk_G$-width at most $k$ or confirms that the
  $\bF$-rank-width of $G$ is larger than $k$.
\end{thm}

\begin{pf*}{Proof.} Let $k$ be fixed and let $\cA$ be the algorithm
  constructed in Theorem \ref{thm:3.3} for $2k+1$. Let $G$ be a
  $\sigma$-symmetric $\bF^*$-graph. We run the algorithm $\cA$ with
  input $(\cM_G,\Pi(G))$.  If it confirms that
  $\bwd{\cM_G,\Pi(G)}>2k+1$, then the $\bF$-rank-width of $G$ is
  greater than $k$ (Corollary \ref{cor:3.1}). If it outputs a
  layout of $\Pi(G)$ of $\lambda_{\cM_G}^{\Pi(G)}$-width at most
  $2k+1$, we can transform it into a layout of $V_G$ of
  $\fcutrk_G$-width at most $k$ by Corollary \ref{cor:3.2}. The fact
  that the algorithm $\cA$ runs in cubic-time concludes the proof.\qed
\end{pf*}

\subsection{Specialisations to Graphs}\label{subsec:3.4}

We specialise in this section the $\bF$-rank-width to directed and oriented
graphs. As we already said, for undirected graphs the $\field{2}$-rank-width 
matches with the rank-width.

\paragraph*{Directed Graphs.}
We recall that the adjacency matrix of a directed graph $G$ is
the $(V_G,V_G)$-matrix $\matg$ over $\field{2}$ where
$\matgind{x}{y}:=1$ if and only if $(x,y)\in E_G$. This matrix is not
symmetric except when $G$ is undirected. In particular,
$\rk(\matgind{X}{V_G\backslash X})$ is \emph{a priori} different from
$\rk(\matgind{V_G\backslash X}{X})$. The quest for finding another
representation of directed graphs by matrices where
$\rk(\matgind{X}{V_G\backslash X}) = \rk(\matgind{V_G\backslash
  X}{X})$ motivates the definition of sigma-symmetry. We now give this
representation.

We recall that $\field{4}$ is the finite field of order four. We let
$\{0,1,\gfa,\gfb\}$ be its elements with the property that
$1+\gfa+\gfb=0$ and $\gfa^3=1$. Moreover, it is of characteristic
$2$. We let $\sigma_4:\field{4}\to \field{4}$ be the automorphism where
$\sigma_4(\gfa) = \gfb$ and $\sigma_4(\gfb) = \gfa$.  It is clearly a
sesqui-morphism. 


For every directed graph $G$, let $\tG:=(V_G,E_G\cup\{(y,x)\vert (x,y)\in E_G\},\ell_G)$ be the
$\gfq^*$-graph where for every pair of vertices $(x,y)$:
\begin{align*}
  \ell_G((x,y)) &:= \begin{cases} 1 & \textrm{if $(x,y)\in
      E_G\ \textrm{and}\ (y,x)\in E_G$},\\ \gfa &
    \textrm{$(x,y)\in E_G\ \textrm{and}\ (y,x)\notin E_G$},\\ \gfb
    & \textrm{$(y,x)\in E_G\ \textrm{and}\ (x,y)\notin E_G$},\\ 0
    & \textrm{otherwise}.
  \end{cases}
\end{align*}

It is straightforward to verify that $\tG$ is $\sigma_4$-symmetric and
is actually the one constructed in Section \ref{sec:3}.  We define the
rank-width of a directed graph $G$, denoted by $\Qrwd{G}$, as the
$\gfq$-rank-width of $\tG$.

\begin{rem}\label{rem:3.2} Let $G$ be an undirected graph. We denote 
  by $\overrightarrow{G}$ the directed graph obtained from $G$ by
  replacing each edge $xy$ in $G$ by two opposite. By the definition of
  $\overrightarrow{G}$ we have $A_G=\mat{\overrightarrow{G}}$. Then
  $\Qrwd{\overrightarrow{G}}= \rwd{G}$ since $\gfq$ is an extension of
  $\field{2}$.
\end{rem}

\begin{figure}[h!]
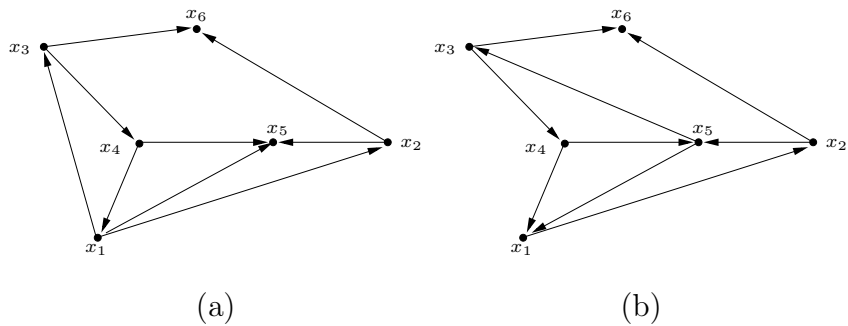

  \centering
  \begin{tabular}{cc}
    \input{./digraph2.pstex_t} &
    \input{./dvminor.pstex_t}\\ (a) &
     (b)
  \end{tabular}
  \caption{(a) A directed graph $G$. (b) The directed graph $\lco{G}{x}_4$.}
  \label{fig:3.1}
\end{figure}

\begin{figure}
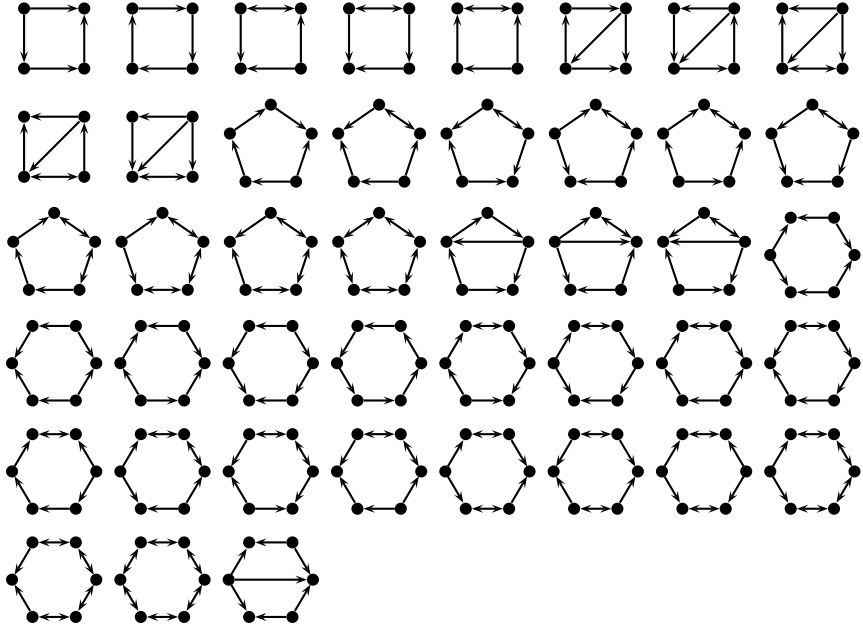

\noindent
\centering
\include{exclp4}
\caption{~Pivot-minor exclusions for directed graphs of $\gfq$-rank-width $1$.}
\label{fig:exclp4}
\end{figure}

We now specialise the notion of vertex-minor. We recall that given a
sesqui-morphism $\sigma:\bF\to \bF$, an element $\lambda$ of $\bF^*$ is said
$\sigma$-compatible if $\sigma(\lambda) = \lambda\cdot \sigma(1)^2$. Since
$\sigma_4(1)=1$, $1$ is $\sigma_4$-compatible and is the only one. We then denote $G*v=G*(v,1)$, and say that a directed graph $H$ is a \emph{vertex-minor} of a directed graph $G$ if $\widetilde{H}$ is a vertex-minor of $\tG$.  One easily
verifies that if a directed graph $H$ is obtained from a directed
graph $G$ by applying a $1$-local-complementation at $x$, then $H$ is
obtained from $G$ by modifying the subgraph induced on the neighbours
of $x$ as shown on Table \ref{tab:3.1}. Figure \ref{fig:3.1} gives an example of
a $1$-local complementation.
In Figure~\ref{fig:exclv4} (resp. Figure~\ref{fig:exclp4}), we give a set of obstructions for directed graphs of $\field{4}$-rank-width $1$ with respect to vertex-minor relation (resp. pivot-minor relation). 

\begin{table}[!h]
  \centering
  \begin{tabular}{cc}
    \begin{tabular}{|c|c|}
      \hline $G$ & $\lco{G}{x}$\\ \hline $\vz\perp \vt$ & $\vz\leftrightarrow
      \vt$\\ \hline $\vz\to \vt$ & $\vz\leftarrow \vt$\\ \hline
      $\vz\leftarrow \vt$ & $\vz\to \vt$\\ \hline $\vz\leftrightarrow
      \vt$ & $\vz\perp \vt$\\ \hline
    \end{tabular} & \qquad\qquad\qquad
    \begin{tabular}{|c|c|}
      \hline $G$ & $\lco{G}{x}$\\ \hline $\vz\perp \vt$ & $\vz\to \vt$\\ \hline
      $\vz\to \vt$ & $\vz\perp \vt$\\ \hline $\vz\leftarrow \vt$ &
      $\vz\leftrightarrow \vt$\\ \hline $\vz\leftrightarrow \vt$ &
      $\vz\leftarrow \vt$\\ \hline
    \end{tabular} \\ (a) &\qquad\qquad\qquad (b) \end{tabular}
  \caption{We use the following notations: $x\to y$ means
    $\ell_G((x,y))=\gfa$, $x\leftarrow y$ means $\ell_G((x,y))=\gfb$,
    $x\leftrightarrow y$ means $\ell_G((x,y))=1$, and $\vz\perp \vt$
    means $\ell_G((x,y)=0)$. \newline (a) Uniform Case: $\vz\leftarrow
    x \to \vt$ or $\vz\to x\leftarrow \vt$ or $\vz\leftrightarrow x
    \leftrightarrow \vt$.\newline (b) Non Uniform Case: $\vz\leftarrow
    x \leftarrow \vt$ or $\vz\to x\leftrightarrow \vt$ or
    $\vz\leftrightarrow x \to \vt$.}
  \label{tab:3.1}
\end{table}

Moreover, as in the undirected case, we have $\pc{G}{xy}=\pc{G}{yx}=\lco{\lco{\lco{G}{x}}{y}}{x}=\lco{\lco{\lco{G}{y}}{x}}{y}$.
As corollaries of Theorem \ref{thm:3.1} and \ref{thm:3.4} we get the
followings. 

\begin{thm}\label{thm:3.5}  For each positive integer $k$, there
  is a finite list $\sC_k$ of directed graphs having at most
  $(6^{k+1}-1)/5$ vertices such that a directed graph $G$ has
  rank-width at most $k$ if and only if no directed graph in $\sC_k$
  is isomorphic to a vertex-minor of $G$.
\end{thm}

\begin{thm}\label{thm:3.6}  For fixed $k$, there exists a cubic-time
  algorithm that, for a directed graph $G$, either outputs a layout of
  $V_G$ of $\cutrk{\gfq}_G$-width at most $k$ or confirms that the
  rank-width of $G$ is larger than $k$.
\end{thm}



\begin{figure}
\noindent
\centering
\include{exclv4}
\caption{~Vertex-minor exclusions for directed graphs of $\gfq$-rank-width $1$.}
\label{fig:exclv4}
\end{figure}

\begin{figure}
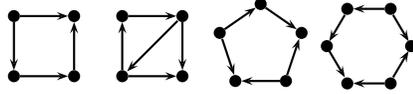

\noindent
\centering
\include{exclp3}
\caption{~Pivot-minor exclusions for oriented graphs of $\field{3}$-rank-width $1$.}
\label{fig:exclp3}
\end{figure}

\paragraph*{Oriented Graphs.}
We can define another parameter in the case of oriented graphs.  Let
$G=(V,A)$ be an oriented graph, and let $\widetilde{G}=(V,E,\ell)$ be
the $\field{3}^*$-graph such that $E=A\cup A'$ where $A'=\{(y,x) |
(x,y)\in A\}$, $\ell((x,y)):=1$ if $(x,y)\in A$ and $\ell((x,y)):=-1$ if
$(x,y)\in A'$. Clearly, $\widetilde{G}$ is a $\sigma$-symmetric
$\field{3}^*$-graph, with $\sigma(x):=-x$. Moreover, one can show
immediately that $\sigma$ is a sesqui-morphism. Note that there is no
$\sigma$-compatible $\lambda$ in $\field{3}^*$, thus no
$\sigma$-local-complementation is defined on $\sigma$-symmetric
$\field{3}^*$-graphs. Nevertheless, oriented graphs of
$\field{3}$-rank-width $k$ are characterized by a finite set of
oriented graphs $\sC_k^{(\field{3},\sigma)}$ of forbidden pivot-minors
(whereas sets $\sC_k^{(\field{4},\sigma)}$ and
${\sC'}_k^{(\field{4},\sigma)}$ contains directed graphs).
In Figure~\ref{fig:exclp3}, we give a set of obstructions for oriented graphs of $\field{3}$-rank-width $1$ with respect to pivot-minor relation.

\begin{figure}[h!]
  \centering
\includegraphics[width=0.5\linewidth]{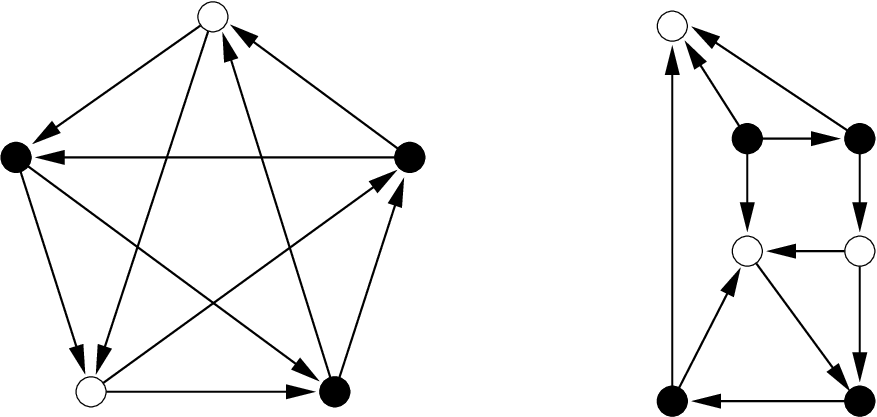}
\caption{Left: an oriented graph of $\field{3}$-rank-width $1$ and $\field{4}$-rank-width $2$ (white/black vertices give a cut of $\field{3}$-rank-width $1$). Right: an oriented graph of $\field{3}$-rank-width $3$ and $\field{4}$-rank-width $2$ (white/black vertices give a cut of $\field{4}$-rank-width $2$).}
\label{fig:orient}
\centering

\end{figure}

$\field{3}$-rank-width and $\field{4}$-rank-width of oriented
graphs are two equivalent parameters, since they are both equivalent
to the clique width. But these two rank parameters are not equal. In
one hand, tournaments of $\field{3}$-rank-width $1$ are exactly
tournaments completely decomposable by \emph{bi-join decomposition}
(see~\cite{KR10b}), and a cut $\{X,Y\}$ in a tournament has
$\field{4}$-rank $1$ if and only if $X$ or $Y$ is a module. Since
there are tournaments completely decomposable by bi-join and prime
w.r.t. the modular decomposition (see~\cite{BHL07}), there are
oriented graphs of $\field{3}$-rank-width $1$ and
$\field{4}$-rank-width at least $2$. On the other hand, the graph on
Figure~\ref{fig:orient} (right) has $\field{4}$-rank-width $2$ and
$\field{3}$-rank-width $3$.

\section{The Second Notion of Rank-Width: $\bF$-Bi-Rank-Width}\label{sec:4}

In Section \ref{subsec:4.1} we define the notion of
\emph{$\bF$-bi-rank-width} for $\bF^*$-graphs, sigma-symmetric or not,
and compare it to clique-width and $\bF$-rank-width. A cubic-time
algorithm for recognising $\bF^*$-graphs of $\bF$-bi-rank-width at
most $k$ is presented in Section \ref{subsec:4.2}. A specialisation to
graphs without colors on edges is given in Section
\ref{subsec:4.3}. 

\subsection{Definitions and Comparisons to Other
  Parameters}\label{subsec:4.1} 

Recall that if $G$ is a $\bF^*$-graph, we denote by $\matg$ the
$(V_G,V_G)$-matrix over $\bF$ where
\begin{align*}
  \matgind{x}{y} &:= \begin{cases} \ell_G((x,y)) & \textrm{if $x\ne y$},\\
    0 & \textrm{otherwise}.  \end{cases}
\end{align*}

As for the notion of $\bF$-rank-width, we use matrix rank functions
for the notion of $\bF$-bi-rank-width.

\begin{defn}[Bi-Cut-Rank Function] For a $\bF^*$-graph  $G$,  we let
  $\fbicutrk_G:2^{V_G}\to \bN$ where $\fbicutrk_G(X) =
  \rk(\matgind{X}{V_G\backslash X}) + \rk(\matgind{V_G\backslash
    X}{X})$ for all $X\subseteq V_G$.
\end{defn}

\begin{lem}\label{lem:4.1} For every $\bF^*$-graph $G$, the function
  $\fbicutrk_G$ is symmetric and submodular.
\end{lem}

\begin{pf*}{Proof.} Let $X$ and $Y$ be subsets
  of $V_G$. We let $A_1:=\matgind{X}{V_G\backslash X}$,
  $A_2:=\matgind{V_G\backslash X}{X}$, $B_1=:\matgind{Y}{V_G\backslash
    Y}$ and $B_2:=\matgind{V_G\backslash Y}{Y}$.  By definition,
  \begin{align*}
    \fbicutrk_G(X) &= \rk(A_1) + \rk(A_2) = \rk(A_2) + \rk(A_1) =
    \fbicutrk_G(V_G\backslash X).
  \end{align*}
  For the submodularity, we have  by definition, 
  \begin{align*}
    \fbicutrk_G(X) + \fbicutrk_G(Y) & = \rk(A_1) + \rk(A_2) + \rk(B_1)
    + \rk(B_2)
  \end{align*}
  By Proposition \ref{prop:3.3},
  {\small 
  \begin{align*}
    \rk(A_1) +\rk(B_1) & \geq  \rk(\matgind{X\cup Y}{V_G\backslash
      X \cap V_G\backslash Y}) + \rk(\matgind{X\cap Y}{V_G\backslash
      X\cup V_G\backslash Y})\\ \intertext{and} \rk(A_2) +\rk(B_2) &
    \geq  \rk(\matgind{V_G\backslash X \cup V_G\backslash Y}{X\cap
      Y}) + \rk(\matgind{V_G\backslash X \cap V_G\backslash Y}{X\cup
      Y}).
  \end{align*}}
  Since $V_G\backslash X\cap V_G\backslash Y = V_G\backslash (X\cup
  Y)$ and $V_G\backslash X \cup V_G\backslash Y= V_G\backslash (X\cap
  Y)$ the second statement holds. \qed
\end{pf*}

\begin{defn}[$\bF$-bi-rank-width] The $\bF$-bi-rank-width of a
  $\bF^*$-graph, denoted by $\fbrwd{G}$, is the $\fbicutrk_G$-width of
  $V_G$.
\end{defn} 

The following proposition compares clique-width and
$\bF$-bi-rank-width when $\bF$ is finite and of order $q$. Its proof is easy.

\begin{prop}\label{prop:4.1} For every $\bF^*$-graph $G$,
  $\frac{1}{2}\fbrwd{G}\leq \cwd{G} \leq 2\cdot q^{\fbrwd{G}} -1$.
\end{prop}

The following compares $\bF$-bi-rank-width and $\bF$-rank-width. 
Let $[G\mapsto \tG]$ be a mapping from $\fieldgr(\bF)$ to $\fieldgr(\bF^2,\widetilde{\sigma})$
such that for every $x,y\in V_G$, $\matind{M_{\tG}}{x}{y}=\gamma \cdot \matgind{x}{y} + \tau \cdot \matgind{y}{x}$ 
for fixed $\gamma,\tau\in (\bF^2)^*$ with $\gamma/\tau \not\in \bF$.
We recall that the mapping constructed in Section \ref{sec:3} respects this property.

\begin{prop}\label{prop:4.2} Let $G$ be a $\bF^*$-graph. Then
  \begin{enumerate}
  \item $\Frwd{\bF^2}{\tG}\leq \fbrwd{G} \leq 4 \cdot
    \Frwd{\bF^2}{\tG}$.

  \item If $G$ is $\sigma$-symmetric for some
    sesqui-morphism $\sigma:\bF\to \bF$, then $\fbrwd{G} = 2\cdot
    \frwd{G}$. 
  \end{enumerate}
\end{prop}

Before proving the proposition, we recall some technical properties
about ranks of matrices. See the following books for more informations
\cite{LIP91,LN97}.

\begin{lem}\label{lem:4.2} 
  \begin{enumerate}
  \item[(i)] Let $M$ be a matrix over $\bF$. If the rank of $M$
    over $\bF$ is $k$, then the rank of $M$ over any finite extension of
    $M$ is $k$.
    
  \item[(ii)] If $A$ and $B$ are two matrices over $\bF$, then
    $\rk(A+B) \leq \rk(A) + \rk(B)$ and $\rk(A\cdot B) \leq
    \min\{\rk(A),\rk(B)\}$. If $a\in \bF^*$, then $\rk(a\cdot A) =
    \rk(A)$.
  \end{enumerate}
\end{lem}

  

By definition of $\tG$, we have: 

\begin{prop} \label{lem:4.3} For every $\bF^*$-graph $G$ and every
  subset $X$ of $V_G$, we have \begin{align*}
    \matind{\mat{\tG}}{X}{V_G\backslash X} = \gamma \cdot
    \matgind{X}{V_G\backslash X} + \tau \cdot
    \matind{\mat{G}^T}{V_G\backslash X}{X}. \end{align*}
\end{prop}


\begin{pf*}{Proof of Proposition \ref{prop:4.2}.}  It is sufficient to
  compare $\cutrk{\bF^2}_{\tG}(X)$ and $\fbicutrk_G(X)$ for every
  subset $X$ of $V_G$.

  \emph{(1)} 
  From Lemma \ref{lem:4.2} and
  Proposition \ref{lem:4.3} we have:
  \begin{align*}
    \rk(\matind{\mat{\tG}}{X}{V_G\backslash X}) & \leq
    \rk(\matgind{X}{V_G\backslash X}) +
    \rk(\matgind{V_G\backslash X}{X}) \\ & =
    \fbicutrk_G(X).
  \end{align*}
  We now prove that $\fbicutrk_G(X) \leq 4\cdot
  \cutrk{\bF^2}_{\tG}(X)$.  Let $M_1:=\matgind{X}{V_G\backslash X}$
  and $M_2:=\matind{\mat{G}^T}{V_G\backslash X}{X}$. We recall that each entry
  of $\mat{\tG}$ is of the form $a\cdot \gamma + b\cdot \tau$ for
  a unique pair $(a,b)\in \bF \times \bF$. Let $\pi_1,\pi_2$ and $\pi_3$ be mappings
  from $\bF^2$ to $\bF$ such that:
  \begin{align*}
    \pi_1(a\cdot \gamma + b\cdot \tau) & = a, \\
    \pi_2(a\cdot \gamma + b\cdot \tau) & = b. 
  \end{align*}
  Clearly,
  $M_1=\pi_1(\matind{\mat{\tG}}{X}{V_G\backslash X})$ and
  $M_2=\pi_2(\matind{\mat{\tG}}{V_G\backslash X}{X})$. It is also
  straightforward to verify that $\pi_1$ and $\pi_2$ are
  homomorphism with respect to the addition. 
  Moreover, for every $c\in \bF$, $\delta\in\bF^2$ and $i\in\{1,2\}$, $\pi_i(c\cdot \delta)=c \cdot \pi_i(\delta)$.

  We let $v_1,\ldots,v_k$ be a column-basis of
  $\matind{\mat{\tG}}{X}{V_G\backslash X}$.  Then for each column-vector
  $v$ in $\matind{\mat{\tG}}{X}{V_G\backslash X}$, 
  $v = \sum_{i\leq k} \alpha_i\cdot v_i$ where $\alpha_i\in
  \bF^2$. Then 
  we have for $j\in
  \{1,2\}$,
  \begin{align*}
    \pi_j(v) & = \sum_{i\leq k} \pi_j(\alpha_i\cdot v_i)\\
    & = \sum_{i\leq k} \pi_j \left (\alpha_i \cdot (\pi_1(v_i)\cdot \gamma + \pi_2(v_i)\cdot\tau) \right)\\
    & = \sum_{i\leq k} \pi_j \left (\alpha_i\cdot\gamma \cdot \pi_1(v_i) + \alpha_i\cdot\tau \cdot \pi_2(v_i) \right)\\
    & = \sum_{i\leq k} \pi_j(\alpha_i\cdot\gamma) \cdot \pi_1(v_i) + \pi_j(\alpha_i\cdot\tau) \cdot \pi_2(v_i)
  \end{align*}
  Thus, every column-vector of $M_j$ is a linear combination of $2k$
 vectors $\pi_1(v_i)$ and $\pi_2(v_i)$ for $i\in \{1,\ldots, k\}$,
  \emph{i.e.} $\rk(M_j) \leq 2k$. Therefore, $\fbicutrk_G(X)
  = \rk(M_1) + \rk(M_2) \leq 4\cdot \cutrk{\bF^2}_{\tG}(X)$.\newline
  
  \emph{(2)} Assume now that $G$ is $\sigma$-symmetric. By definition
  of $\fbicutrk_G$, we have $\fbicutrk_G(X) =
  \rk(\matgind{X}{V_G\backslash X})+\rk (\matgind{V_G\backslash
    X}{X})$. But since $G$ is $\sigma$-symmetric, by Lemma
  \ref{lem:3.1}, we have $\rk(\matgind{X}{V_G\backslash X}) =
  \rk(\matgind{V_G\backslash X}{X})$. We can then conclude that
  $\fbicutrk_G(X) = 2\cdot \fcutrk_G(X)$.\qed
\end{pf*}

The notion of local complementation defined in Section
\ref{subsec:3.2} also preserves the $\bF$-bi-rank-width. 

\begin{lem}\label{lem:4.4} Let $G$ be a $\bF^*$-graph and $\lambda$
  an element in $\bF^*$. If $\lc{G}{x}$ is the $\lambda$-local complementation
  of $G$ at $x$, then for every subset $X$ of $V_G$, we have
  $\fbicutrk_{\lc{G}{x}}(X) = \fbicutrk_G(X)$.
\end{lem}

\begin{pf*}{Proof.} Assume by symmetry that $x$ is in $X$. Let $y$ be
  a neighbor of $x$ in $X$. If we apply a $\lambda$-local
  complementation at $x$, we obtain
  $\matind{\mat{\lc{G}{x}}}{y}{V_G\backslash X}$ by adding $\lambda\cdot
  \matgind{y}{x}\cdot \matgind{x}{V_G\backslash X}$ to
  $\matgind{y}{V_G\backslash X}$.  Therefore,
  $\rk(\matind{\mat{\lc{G}{x}}}{X}{V_G\backslash X}) =
  \rk(\matgind{X}{V_G\backslash X})$. Similarly, we obtain
  $\matind{\mat{\lc{G}{x}}}{V_G\backslash X}{y}$ by adding to
  $\matgind{V_G\backslash X}{y}$ the column $\lambda \cdot
  \matgind{V_G\backslash X}{x}\cdot \matgind{x}{y}$. Again,
  $\rk(\matind{\mat{\lc{G}{x}}}{V_G\backslash X}{X}) =
  \rk(\matgind{V_G\backslash X}{X})$. We can thus conclude that
  $\fbicutrk_{\lc{G}{x}}(X) = \fbicutrk_G(X)$. \qed
\end{pf*}

As a corollary, we get the following.

\begin{cor}\label{cor:4.1} Let $G$ and $H$ be two $\bF^*$-graphs. If
  $H$ is locally equivalent to $G$, then the $\bF$-bi-rank-width of
  $H$ is equal to the $\bF$-bi-rank-width of $G$. If $H$ is a
  vertex-minor of $G$, then the $\bF$-bi-rank-width of $H$ is at most
  the $\bF$-bi-rank-width of $G$.
\end{cor}

Note that the pivot-complementation in Section \ref{subsec:3.2} is not well defined in the case of no-sigma-symmetric graphs.
Currently, we do not have a characterisation of $\bF^*$-graphs of
bounded $\bF$-bi-rank-width as the one in Theorem \ref{thm:3.1}. We
leave it as an open question. 
Moreover, this notion of vertex-minor is not a
well-quasi-order on $\bF^*$-graphs of bounded $\bF$-bi-rank-width (see
Remark \ref{rem:4.1}).

\subsection{Recognizing $\bF$-Bi-Rank-Width at Most $k$}\label{subsec:4.2}

We will give here, for fixed $k$ and a fixed finite field 
$\bF$, a cubic-time algorithm that decides
whether a $\bF^*$-graph has $\bF$-bi-rank-width at most $k$. The
algorithm is in the same spirit as the one in Section
\ref{subsec:3.3}.

We recall that if $G$ is a $\bF^*$-graph, we denote by $(\cM_G,\Pi(G))$
the partitioned matroid represented over $\bF$ where $\cM_G$ is the
matroid represented by the $(V_G,V_G\cup V_G')$-matrix over $\bF$
($V_G'$ is an isomorphic copy of $V_G$) 
\begin{align*}
  \begin{matrix}
    & V_G\qquad V_G'\\
    V_G & \Big(\begin{matrix}  I_{|V_G|} \qquad  \matg\end{matrix}\Big)
  \end{matrix}
\end{align*} and
$\Pi(G) := \{P_x\mid x\in V_G\}$ with $P_x:=\{x,x'\}$. 

As corollaries of Proposition \ref{prop:3.7} we get the followings. 

\begin{cor}\label{cor:4.2} Let $G$ be a $\bF^*$-graph. For every
  $X\subseteq V_G,\ \lambda_{\cM_G}^{\Pi(G)}(P) = \fbicutrk_G(X)
  +1$ where $P:=\{P_x~|~x\in X\}$.
\end{cor}

\begin{cor}\label{cor:4.3} Let $G$ be a 
  $\bF^*$-graph and let $p:V_G\to \Pi(G)$ be the bijective function
  such that $p(x)=P_x$. If $(T,\cL)$ is a layout of $\Pi(G)$ of
  $\lambda_{\cM_G}^{\Pi(G)}$-width $k+1$, then $(T,\cL\circ p)$ is a
  layout of $V_G$ of $\fbicutrk_G$-width $k$. Conversely, if $(T,\cL)$
  is a layout of $V_G$ of $\fbicutrk_G$-width $k$, then $(T,\cL\circ
  p^{-1})$ is a layout of $\Pi(G)$ of $\lambda_{\cM_G}^{\Pi(G)}$-width
  $k+1$.
\end{cor}

\begin{thm}[Checking $\bF$-Bi-Rank-Width at most $k$]\label{thm:4.1}
  For a fixed finite field $\bF$ and a fixed integer $k$, 
  there exists a cubic-time algorithm that, for a
  $\bF^*$-graph $G$, either outputs a layout of $V_G$ of
  $\fbicutrk_G$-width at most $k$ or confirms that the
  $\bF$-bi-rank-width of $G$ is larger than $k$.
\end{thm}

\begin{pf*}{Proof.} Let $k$ be fixed and let $\cA$ be the algorithm
  constructed in Theorem \ref{thm:3.3} for $k+1$. Let $G$ be a
  $\bF^*$-graph. We run the algorithm $\cA$ with input
  $(\cM_G,\Pi(G))$.  If it confirms that $\bwd{\cM_G,\Pi(G)}>k+1$,
  then the $\bF$-bi-rank-width of $G$ is greater than $k$ (Corollary
  \ref{cor:4.2}). If it outputs a layout of $\Pi(G)$ of
  $\lambda_{\cM_G}^{\Pi(G)}$-width at most $k+1$, we can transform it
  into a layout of $V_G$ of $\fbicutrk_G$-width at most $k$ by
  Corollary \ref{cor:4.3}. The fact that the algorithm $\cA$ runs in
  cubic-time concludes the proof.\qed
\end{pf*}

\subsection{A Specialisation to Graphs}\label{subsec:4.3}

We now define our second notion of rank-width for directed graphs. We
recall that a graph $G$ is seen, also denoted by $G$, as the
$\bF_2^*$-graph where $\matgind{x}{y}:=1$ if and only if $(x,y)\in
E_G$. The \emph{bi-rank-width} of a graph $G$, denoted by $\brwd{G}$,
it its $\bF_2$-bi-rank-width. It is straightforward to verify that if
$G$ is undirected, \ie, if $E_G$ is symmetric, then $\brwd{G}=2\cdot
\rwd{G}$.

A directed graph is \emph{strongly connected} if for every pair $(x,y)$ of vertices, there is a directed path from $x$ to $y$. Clearly in a strongly connected graph $G$, for every $\emptyset \subsetneq X \subsetneq V_G$, we have $\fbicutrk_G(X)\ge 2$. It is straightforward to verify that strongly connected graphs of bi-rank-width $2$ are exactly the graphs completely decomposable by Cunningham's split decomposition of directed graphs~\cite{Cun82}. 

The $1$-local complementation of a directed graph seen as a
$\bF_2^*$-graph is the one defined by Bouchet \cite{BOU87} and
Fon-Der-Flaass \cite{FON96}. One easily verifies that if $H$ is
obtained by applying a $1$-local complementation at $x$ to $G$, then
$(\vz,\vt)\in E_H$ if and only if:
\begin{itemize}
\item[-]  $(\vz,\vt)\notin E_G$, $(\vz,x)\in E_G$ and $(x,\vt)\in
  E_G$ or,
\item[-] $(\vz,\vt)\in E_G$, and either $(\vz,x)\notin E_G$ or $(x,\vt)\notin
  E_G$.
\end{itemize}

\begin{figure}[h!]
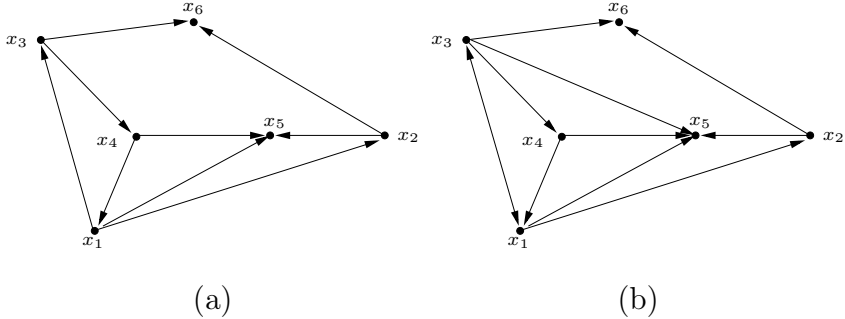

  \centering
  \begin{tabular}{cc}
    \input{./digraph2.pstex_t} &
    \input{./dbminor.pstex_t}\\ (a) &
     (b)
  \end{tabular}
  \caption{(a) A directed graph seen as a $\bF_2^*$-graph. (b) Its $1$-local complementation at
    $x_4$.} 
  \label{fig:4.1}
\end{figure}

Figures \ref{fig:4.1} illustrates a $1$-local complementation of a
directed graph seen as a $\bF_2^*$-graph.

\begin{figure}[h!]
  \centering
  \input{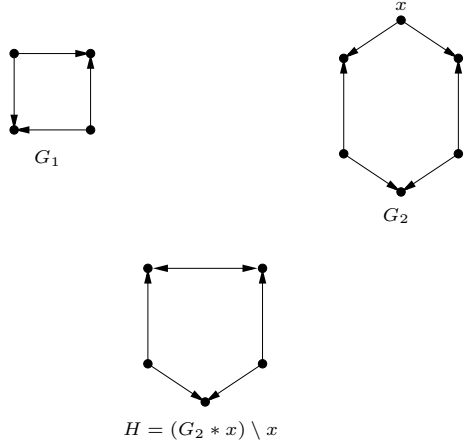}
  \caption{$G_1$ and $G_2$ are graphs in $\mathcal{EC}$. Each graph in
    $\mathcal{EC}$, seen as a $\bF_2^*$-graph, is isomorphic to its
    $1$-local complementations. This is not the case if we consider
    them as a $\sigma_4$-symmetric $\bF_4^*$-graph. For instance, $H$
    is a vertex-minor of $G_2$ seen as a $\sigma_4$-symmetric
    $\bF_4^*$-graph.}
  \label{fig:4.2}
\end{figure}

\begin{figure}[h!]
  \centering
  \input{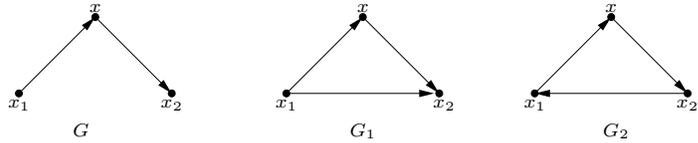}
  \caption{$G_1$ is a vertex-minor of $G$ seen as a $\bF_2^*$-graph
    and the only one locally equivalent to it, and $G_2$ is a
    vertex-minor of $G$ seen as a $\sigma_4$-symmetric
    $\bF_4^*$-graph. $G_1$ is not isomorphic to $G_2$.}
  \label{fig:4.3}
\end{figure}

We notice that the $1$-local complementation of a directed graph seen
as a $\bF_2^*$-graph can be different from the one when we consider it
as a $\sigma_4$-symmetric $\bF_4^*$-graph (see Section
\ref{subsec:3.4}). Figures \ref{fig:4.2} and \ref{fig:4.3} illustrate
this observation. We leave open the question of finding a notion of
vertex-minor for directed graphs, that not only lets invariant
$\bF_4$-rank-width and $\bF_2$-bi-rank-width, but also is independent
of the representation.

\begin{rem}\label{rem:4.1} Directed graphs of bounded bi-rank-width
  are not well-quasi-ordered by the vertex-minor relation. In fact
  the class $\mathcal{EC}$ of directed even cycles such that each vertex has
  either in-degree $2$ or out-degree $2$ are of bounded bi-rank-width
  and are not well-quasi-ordered by vertex-minor relation since none
  of them is a vertex-minor of another. In fact each of them is
  isomorphic to its $1$-local complementations. Figure \ref{fig:4.2}
  illustrates such cycles.
\end{rem}

\section{Algebraic Operations for $\bF$-Rank-Width and
  $\bF$-Bi-Rank-Width}\label{sec:5} 

Courcelle and the first author gave in \cite{CK09} graph operations
that characterise exactly the notion of rank-width of undirected
graphs. These operations are interesting because they allow to check
monadic second-order properties on undirected graph classes of bounded
rank-width without using clique-width operations. This important in a
practical point of view since it allows to decrease by $1$ the hidden
towers of exponents due to the generality of the method. We give in
Section \ref{subsec:5.1} graph operations, that generalise the ones in
\cite{CK09} and that characterise exactly $\bF$-rank-width. A
specialisation that allows to characterise exactly $\bF$-bi-rank-width
is then presented in Section \ref{subsec:5.2}. 

We let $\bF$ be a fixed finite field along this section.  For a fixed
positive integer $k$, we let $\bF^k$ be the set of row vectors of
length $k$.

\subsection{Operations Characterising
  $\bF$-Rank-Width} \label{subsec:5.1} 

The operations are easy adaptations of the ones in \cite{CK09}. We let
$\sigma:\bF\to \bF$ be a fixed sesqui-morphism. If $u:=(u_1,\ldots,u_k)\in
\bF^k$, we let $\sigma(u)$ be $(\sigma(u_1), \ldots,
\sigma(u_k))$. Similarly, if $M=(m_{i,j})$ is a matrix, we let
$\sigma(M)$ be the matrix $(\sigma(m_{i,j}))$. In this section we deal
with $\sigma$-symmetric $\bF^*$-graphs. So, we will say graph instead
of $\sigma$-symmetric $\bF^*$-graph. 

An \emph{$\bF^{k}$-coloring} of a graph $G$ is a mapping
$\gamma_G:V_G\to \bF^{k}$ with no constraint on the values of $\gamma$
for adjacent vertices. An \emph{$\bF^{k}$-colored graph} $G$ is a
tuple $(V_G,E_G,\ell_G,\gamma_G)$ where $(V_G,E_G,\ell_G)$ is a graph
and $\gamma_G$ is an $\bF^{k}$-coloring of $(V_G,E_G,\ell_G)$. Notice
that an $\bF^k$-colored graph has not only its edges colored with
colors from $\bF$, but also its vertices with colors from
$\bF^k$. With an $\bF^k$-colored graph $G$, we associate the
$(V_G\times [k])$-matrix $\Gamma_G$, the row vectors of which are the
vectors $\gamma_G(x)$ in $\bF^{k}$ for $x$ in $V_G$.

The following is a binary graph operation that combines several
operations consisting in adding colored edges between its disjoint
arguments and recolor them independently.

\begin{defn}[Bilinear Products]\label{defn:5.1}  Let $k,\ell$ and $m$
  be positive integers and let $M,N$ and $P$ be $k\times
  \ell,\ k\times m$ and $\ell\times m$ matrices, respectively, over
  $\bF$. For an $\bF^k$-colored graph $G$ and an $\bF^{\ell}$-colored
  graph $H$, we let $G\otimes_{M,N,P} H$ be the $\bF^m$-colored graph
  $K:=(V_G\cup V_H, E_G\cup E_H\cup E', \ell_K, \gamma_K)$
  where:
  \begin{align*}
    E' &:= \{xy\mid x\in V_G,\ y\in
    V_H\ \textrm{and}\ \gamma_G(x)\cdot M\cdot \sigma(\gamma_H(y))^T
    \ne 0\},\\ 
    \ell_K((x,y)) &:= 
    \begin{cases} 
      \ell_G((x,y)) & \textrm{if $x,y\in V_G$},\\ 
      \ell_H((x,y)) & \textrm{if $x,y\in V_H$},\\ 
      \gamma_G(x)\cdot M\cdot\sigma(\gamma_H(y))^T & \textrm{if $x\in
        V_G$, $y\in V_H$},\\
     \sigma\left(\gamma_G(y)\cdot M\cdot\sigma(\gamma_H(x))^T\right) & \textrm{if $y\in V_G$, $x\in V_H$}. 
    \end{cases} \\
    \gamma_K(x) & := \begin{cases} \gamma_G(x)\cdot N &
      \textrm{if}~x\in V_G,\\ \gamma_H(x)\cdot P & \textrm{if}~x\in
      V_H. 
    \end{cases}
  \end{align*}
\end{defn}

\begin{defn}[Constants]\label{defn:5.2} For each $u\in \bF^{k}$, we
  let $\const{u}$ be a constant denoting a $\bF^k$-colored graph with
  one vertex colored by $u$ and no edge.
\end{defn}

We denote by $\cC_n^{\bF}$ the set $\{\const{u}~|~u\in \bF^{1}\cup
\cdots\cup \bF^{n}\}$. We let $\cR_n^{(\bF,\sigma)}$ be the set of
bilinear products $\otimes_{M,N,P}$ where $M,N$ and $P$ are
respectively $k\times \ell$, $k\times m$ and $\ell \times m$ matrices
for $k,\ell, m \leq n$. Each term $t$ in
$T(\cR_n^{(\bF,\sigma)},\cC_n^{\bF})$ defines, up to
isomorphism, a $\sigma$-symmetric $\bF^*$-graph $val(t)$. We
write by abuse of notation $G=val(t)$ to say that $G$ is
isomorphic to $val(t)$. 

One easily verifies that the operations $\otimes_{M,N,P}$ can be
defined in terms of the disjoint union and quantifier-free
operations. The following is thus a corollary of results in
\cite{CER93,CM02}.

\begin{thm}\label{thm:5.1} For each monadic second-order property
  $\varphi$, there exists an algorithm that checks for every term
  $t\in T(\cR_n^{(\bF,\sigma)},\cC_n^{\bF})$, in time $O(|t|)$, if the
  $\sigma$-symmetric $\bF^*$-graph defined by this term, up to
  isomorphism, satisfies $\varphi$.
\end{thm}

The principal result of this section is the following.

\begin{thm}\label{thm:5.2} A graph $G$ has $\bF$-rank-width at most
  $n$ if and only if it is isomorphic to $val(t)$ for some term $t$ in
  $T(\cR_n^{(\bF,\sigma)},\cC_n^{\bF})$.
\end{thm}

Let $\sigma_1:\field{2}\to \field{2}$ be the identity automorphism. As
a corollary of Theorem \ref{thm:5.2}, we get the following. 

\begin{thm}[\cite{CK09}]\label{thm:5.3} An undirected graph has
  rank-width at most $n$ if and only if it is isomorphic to
  $val(t)$ for some term $t$ in
  $T(\cR_n^{(\field{2},\sigma_1)},\cC_n^{\field{2}})$.
\end{thm}

We can now begin the proof of Theorem \ref{thm:5.2}. It is similar to
the one in \cite{CK09}. 

\begin{lem}\label{lem:5.1} If $K=G\otimes_{M,N,P} H$, then
  $\matind{\mat{K}}{V_G}{V_H} = \Gamma_G\cdot M\cdot
  \sigma(\Gamma_H)^T$ and $\Gamma_K=\left(\begin{smallmatrix}
    \Gamma_G\cdot N\\ \Gamma_H\cdot
    P \end{smallmatrix}\right)$. Moreover, $K$ is isomorphic to
  $H\otimes_{M',P,N} G$ where $M'=\frac{1}{\sigma(1)^2}\cdot
  \sigma(M)^T$.
\end{lem} 

\begin{lem}\label{lem:5.2} Let $t=c\bullet t'$ where $t'\in
  T(\cR_n^{(\bF,\sigma)},\cC_n^{\bF})$ and $c\in
  Cxt(\cR_n^{(\bF,\sigma)},\cC_n^{\bF})\backslash Id$. If $G=val(t)$
  and $H=val(t')$, then 
  \begin{align*}
    \matgind{V_H}{V_G\backslash V_H} & = \Gamma_H\cdot B,\\
    \Gamma_{G[V_H]} &= \Gamma_H\cdot C.
  \end{align*}
  for some matrices $B$ and $C$.
\end{lem}

\begin{pf*}{Proof.}  We prove it by induction on the structure of
  $c$. We identify two cases (the two other cases are similar by
  symmetry and Lemma \ref{lem:5.1}).  

  \begin{description}
  \item[Case 1] $c=id\otimes_{M,N,P} t''$. Then, $G=H\otimes_{M,N,P} K$ where
    $K=val(t'')$.  By Lemma \ref{lem:5.1},
    \begin{align*}
      \matgind{V_H}{V_G\backslash V_H} &= \Gamma_H\cdot M\cdot
      \sigma(\Gamma_K)^T,\\
      \Gamma_{G[H]} & = \Gamma_H\cdot N.
    \end{align*}
    We let $B=M\cdot \sigma(\Gamma_K)^T$ and $C=N$. 

  \item[Case 2] $c=c'\otimes_{M,N,P} t''$. We let $G'=val(c'\bullet t')$ and
    $K=val(t'')$. Hence, $G=G'\otimes_{M,N,P} K$. By definition and
    Lemma \ref{lem:5.1},
    \begin{align*} 
      \matgind{V_H}{V_G\backslash V_H} &= \begin{pmatrix}
        \matind{\mat{G'}}{V_H}{V_{G'}\backslash V_H} &&&&
        \matind{(\Gamma_{G'}\cdot M\cdot
          \sigma(\Gamma_K)^T)}{V_H}{V_K} \end{pmatrix}
    \end{align*}
    By inductive hypothesis, $\matind{\mat{G'}}{V_H}{V_{G'}\backslash
      V_H} = \Gamma_H\cdot B'$ for some matrix $B'$. Moreover,
    $\matind{(\Gamma_{G'}\cdot M\cdot \sigma(\Gamma_K)^T)}{V_H}{V_K}
    = \Gamma_{G'[V_H]}\cdot M\cdot \sigma(\Gamma_K)^T$. But by
    inductive hypothesis, $\Gamma_{G'[V_H]}=\Gamma_H\cdot C'$ for some
    matrix $C'$. Then, $\matgind{V_H}{V_G\backslash V_H} =
    \Gamma_H\cdot B$ where $B=\left(\begin{matrix} B' &&&& C'\cdot M\cdot
      \sigma(\Gamma_K)^T\end{matrix}\right)$. Moreover, $\Gamma_{G[H]}
      = \Gamma_H\cdot C$ where $C=C'\cdot N$ since $\Gamma_{G[V_H]} =
      \Gamma_{G'[V_H]}\cdot N$. \qed
  \end{description}
\end{pf*}

We now prove the ``if direction'' of Theorem \ref{thm:5.2} in the
following. 

\begin{prop}\label{prop:5.1} If $G$ is isomorphic to $val(t)$ for a
  term $t$ in $T(\cR_n^{(\bF,\sigma)},\cC_n^{\bF})$, then $\frwd{G}
  \leq n$.
\end{prop}

\begin{pf*}{Proof.} Let $T$ be the syntactic tree of $t$. By
  definition, there exists a bijective function $\cL:V_G\to L_T$ where
  $L_T$ is the set of leaves of $t$, then of $T$. We let $(T,\cL)$ be
  a layout of $V_G$. In order to prove that the $\fcutrk_G$-width of
  $(T,\cL)$ is at most $n$, it is sufficient to prove that for each
  subgraph $H$ of $G$ associated to a sub-term $t'$ of $t$,
  $\fcutrk_G(V_H)\leq n$. However, we have proved in Lemma
  \ref{lem:5.2} that $\matgind{V_H}{V_G\backslash V_H} = \Gamma_H\cdot
  B$ for some matrix $B$. And since each such $H$ is $\bF^k$-colored
  for some $k\leq n$, we are done.\qed
\end{pf*}

The following proves the ``only if direction'' of Theorem
\ref{thm:5.2}. 

\begin{prop}\label{prop:5.2} If $\frwd{G}\leq n$, then $G$ is
  isomorphic to $val(t)$ for a term $t$ in
  $T(\cR_n^{(\bF,\sigma)},\cC_n^{\bF})$.
\end{prop}

Let us first introduce another notion. Let $V$ be a subset of $V_G$. A
subset $X$ of $V$ is called a \emph{vertex-basis of
  $\matgind{V}{V_G\backslash V}$} if $\{\matgind{x}{V_G\backslash
  V}\mid x\in X\}$ is linearly independent and generates the row space
of $\matgind{V}{V_G\backslash V}$.

\begin{pf*}{Proof.} Assume first that $G$ is connected. Let $(T,\cL)$
  be a layout of $V_G$ of $\fcutrk_G$-width at most $n$. We pick an
  edge of $T$, subdivide it and root the new tree $T'$ by considering
  the new node as the root. For each node $u$ of $T'$, we let $G_u$ be
  the subgraph of $G$ induced by the vertices that are in correspondence
  with the leaves of the sub-tree of $T'$ rooted at $u$. We let $r(u)$
  be $\fcutrk_G(V_{G_u})$.

  \begin{lem}\label{lem:5.3} For each node $u$ of $T'$, we can
    construct a term $t_u$ in $T(\cR_n^{(\bF,\sigma)},\cC_n^{\bF})$
    such that $val(t_u)$ is isomorphic to $G_u$ and is a
    $\bF^{r(u)}$-colored graph. There exists moreover a vertex-basis
    $X_u$ of $\matgind{V_{G_u}}{V_G\backslash V_{G_u}}$ such that
    $\matgind{V_{G_u}}{V_G\backslash V_{G_u}} = \Gamma_{val(t_u)}\cdot
    \matgind{X_u}{V_G\backslash V_{G_u}}$.
  \end{lem}

  It is clear that if $r$ is the root of $T'$, then $G=val(t_r)$ where
  $t_r$ is the term in $T(\cR_n^{(\bF,\sigma)},\cC_n^{\bF})$
  constructed in Lemma \ref{lem:5.3}.

  Assume now that $G$ is not connected and let $G_1,\ldots, G_m$ be
  the connected components of $G$. By Lemma \ref{lem:5.3}, we can
  construct terms $t_1,\ldots, t_m$ that defines, up to
  isomorphism, respectively, $G_1,\ldots, G_m$. It is clear
  that $((\ldots((t_1\otimes_{O,O,O} t_2) \otimes_{O,O,O}
  t_3)\ldots) \otimes_{O,O,O} t_m)$ is isomorphic to $G$ where
  $O$ is the null matrix of order $1\times 1$. This concludes the proof of the
  proposition and therefore of Theorem \ref{thm:5.2}. \qed

  \begin{pf*}{Proof of Lemma \ref{lem:5.3}.} We prove it by
    induction on the number of vertices of $G_u$. Let $u$ be a node in
    $T'$. 

    If $G_u$ is a single vertex $x$, then since $G$ is connected we let
    $t_u:= \const{1}$ and $X_u:=\{x\}$. It is clear that $t_u$ and
    $X_u$ verify the statements of the lemma. 

    Assume now that $G_u$ has at least two vertices. Then $u$ has two
    sons $u_1$ and $u_2$ so that $G_{u_1}$ and $G_{u_2}$ have less
    vertices than $G_u$. By inductive hypothesis, there exist
    $t_{u_i}$ and $X_{u_i}$, for $i=1,2$, verifying the statements of
    the lemma.  We let $r(u_1):=h$ and $r(u_2):=k$. We let $X_{u_1} :=
    \{x_1,\ldots, x_h\}$ and $X_{u_2} := \{y_1, \ldots, y_k\}$. We let
    $M:=\frac{1}{\sigma(1)}\cdot \matgind{X_{u_1}}{X_{u_2}}$, and
    $H=val(t_{u_1})$ and $K=val(t_{u_2})$.

    \begin{claim}\label{claim:5.1} $\matgind{V_{G_{u_1}}}{V_{G_{u_2}}} =
      \Gamma_{H} \cdot M \cdot \sigma(\Gamma_{K})^T$. 
    \end{claim}

    \begin{pf*}{Proof of Claim \ref{claim:5.1}.} Let $x\in
      V_{G_{u_1}}$ and $y\in V_{G_{u_2}}$. By inductive hypothesis, $
      \matgind{x}{V_G\backslash V_{G_{u_1}}} = \gamma_{H}(x)\cdot
      \matgind{X_{u_1}}{V_G\backslash V_{G_{u_1}}}$ and
      $\matgind{y}{V_G\backslash V_{G_{u_2}}} =\gamma_K(y)\cdot
      \matgind{X_{u_2}}{V_G\backslash V_{G_{u_2}}}$. Hence,
      $\gamma_{H}(x)\cdot M = \frac{1}{\sigma(1)} \cdot
      \matgind{x}{X_{u_2}}$. Therefore,
      \begin{align*}
        \gamma_{H}(x)\cdot M\cdot \sigma(\gamma_{K}(y))^T
        &= \frac{1}{\sigma(1)}\cdot \matgind{x}{X_{u_2}} \cdot
        \sigma(\gamma_{K}(y))^T\\ &= \frac{1}{\sigma(1)}\cdot
        \sigma(\matgind{X_{u_2}}{x})^T \cdot
        \sigma(\gamma_{K}(y))^T\\ &= \frac{1}{\sigma(1)}\cdot
        \sigma(\gamma_{K}(y)) \cdot
        \sigma(\matgind{X_{u_2}}{x})\\ &=
        \sigma(\gamma_{K}(y)\cdot \matgind{X_{u_2}}{x})\\ &=
        \sigma(\matgind{y}{x}) 
           = \matgind{x}{y}.\qed
      \end{align*}
    \end{pf*}

    It remains now to find a vertex-basis $X_u$ of
    $\matgind{V_{G_u}}{V_G\backslash V_{G_u}}$ and matrices $N$ and
    $P$ such that $\matgind{V_{G_u}}{V_G\backslash V_{G_u}} =
    \left(\begin{smallmatrix} \Gamma_H\cdot N\\ \Gamma_K\cdot
      P\end{smallmatrix} \right) \cdot \matgind{X_u}{V_G\backslash
        V_{G_u}}$.

      It is straightforward to verify that
      $\{\matgind{z}{V_G\backslash V_{G_u}}\mid z\in X_{u_1}\cup
      X_{u_2}\}$ generates the row space of
      $\matgind{V_{G_u}}{V_G\backslash V_{G_u}}$. Therefore, we can
      find a vertex-basis $X_u$ of $\matgind{V_{G_u}}{V_G\backslash
        V_{G_u}}$ which is a subset $X_{u_1}\cup X_{u_2}$. That means,
      for each $z\in X_{u_1}\cup X_{u_2}$, there exists a row vector
      $b_z$ such that $\matgind{z}{V_G\backslash V_{G_u}} = b_z\cdot
      \matgind{X_u}{V_G\backslash V_{G_u}}$. We let $t_u = t_{u_1}
      \otimes_{M,N,P} t_{u_2}$ where:
      \begin{align*}
        N &:= \begin{pmatrix} b_{x_1} & \cdots & b_{x_h} \end{pmatrix}^T & P
        &:= \begin{pmatrix} b_{y_1} & \cdots & b_{y_h} \end{pmatrix} ^T\\
      \end{align*}
      From Claim \ref{claim:5.1} it remains to show that
      $\Gamma_H\cdot N \cdot \matgind{X_u}{V_G\backslash V_{G_u}} =
      \matgind{V_{G_{u_1}}}{V_G\backslash V_{G_u}}$ and $\Gamma_K\cdot
      P \cdot \matgind{X_u}{V_G\backslash V_{G_u}} =
      \matgind{V_{G_{u_2}}}{V_G\backslash V_{G_u}}$. For that it is
      sufficient to prove, for each $t$ in $V_G\backslash V_{G_u}$,
      that $\matgind{X_{u_1}}{t} = N\cdot \matgind{X_u}{t}$ and
      $\matgind{X_{u_2}}{t} = P\cdot \matgind{X_u}{t}$. But, this is a
      straightforward computation by the definitions of $N$, $P$ and
      $X_u$. This concludes the proof of the lemma. \qed
  \end{pf*}
\end{pf*}

\subsection{Graph Operations for
  $\bF$-Bi-Rank-Width} \label{subsec:5.2}  

In this section we specialise the graph operations in
$\cR_n^{(\bF,\sigma)}$ in order to give graph operations that
characterise exactly $\bF$-bi-rank-width. We start by some
notations. Let $k_1$ and $k_2$ be positive integers. An
\emph{$\bF^{k_1,k_2}$-bi coloring} of a $\bF^*$-graph $G$ is a couple
of mappings $\gamma_G^+:V_G\to \bF^{k_1}$ and $\gamma_G^-:V_G\to
\bF^{k_2}$. An \emph{$\bF^{k_1,k_2}$-bi colored graph} is a tuple
$(V_G,E_G,\ell_G,\gamma_G^+,\gamma_G^-)$ where $(V_G,E_G,\ell_G)$ is a
$\bF^*$-graph and $(\gamma_G^+,\gamma_G^-)$ is a $\bF^{k_1,k_2}$-bi
coloring. With an $\bF^{k_1,k_2}$-bi colored graph $G$ we associate
the $(V_G,[k_1])$ and $(V_G,[k_2])$-matrices $\Gamma_G^+$ and
$\Gamma_G^-$, the row vectors of which are respectively
$\gamma_G^+(x)$ and $\gamma_G^-(x)$ for $x$ in $V_G$. 

\begin{defn}\label{defn:5.3} Let $k_1,k_2,\ell_1,\ell_2,m_1$ and $m_2$
  be positive integers. Let $M_1$, $M_2$, $N_1$, $N_2$, $P_1$ and
  $P_2$ be respectively $k_1\times \ell_1$, $k_2\times \ell_2$,
  $k_1\times m_1$, $k_2\times m_2$, $\ell_1\times m_1$ and
  $\ell_2\times m_2$-matrices.  For a $\bF^{k_1,k_2}$-bi colored graph
  $G$ and a $\bF^{\ell_1,\ell_2}$-bi colored graph $H$, we let
  $G\otimes_{M_1,M_2,N_1,N_2,P_1,P_2} H$ be the $\bF^{m_1,m_2}$-bi colored graph
  $K:=(V_G\cup V_H, E_G\cup E_H\cup E_1\cup E_2, \ell_K,
  \gamma_K^+,\gamma_k^-)$ where:
    \begin{align*}
      E_1 &:= \{(x,y)\mid x\in V_G,\ y\in
      V_H\ \textrm{and}\ \gamma_G^+(x)\cdot M_1 \cdot
      (\gamma_H^-(y))^T\ne 0\},\\ E_2 &:= \{(y,x)\mid x\in V_G,\ y\in
      V_H\ \textrm{and}\ \gamma_G^-(x)\cdot M_2 \cdot
      (\gamma_H^+(y))^T\ne 0\},\\ \ell_K((x,y)) &:= \begin{cases}
        \ell_G((x,y)) & \textrm{if $x,y\in V_G$},\\ \ell_H((x,y)) &
        \textrm{if $x,y\in V_H$},\\ \gamma_G^+(x)\cdot M_1\cdot
        (\gamma_H^-(y))^T & \textrm{if $x\in V_G$ and $y\in V_H$},
        \\ \gamma_G^-(y)\cdot M_2\cdot (\gamma_H^+(x))^T &
        \textrm{if $y\in V_G$ and $x\in V_H$}, \end{cases} \\\gamma_K^+(x) &
      := \begin{cases} \gamma_G^+(x)\cdot N_1 & \textrm{if}~x\in
        V_G,\\ \gamma_H^+(x)\cdot P_1 & \textrm{if}~x\in
        V_H, \end{cases} \\\gamma_K^-(x) & := \begin{cases}
        \gamma_G^-(x)\cdot N_2 & \textrm{if}~x\in
        V_G,\\ \gamma_H^-(x)\cdot P_2 & \textrm{if}~x\in
        V_H, \end{cases}
    \end{align*}
\end{defn}

\begin{defn}\label{defn:5.4} For each pair $(u,v)\in \bF^{k_1}\times
  \bF^{k_2}$, we let $\const{u\cdot v}$ be the constant denoting a
  $\bF^{k_1,k_2}$-bi colored graph with a single vertex and no edge.
\end{defn}

We let $\cBC_n^{\bF}$ be the set $\{\const{u\cdot v}\mid (u,v)\in
\bF^{k_1}\times \bF^{k_2}\ \textrm{and}\ k_1+k_2\leq n\}$.  We denote
by $\cBR_n^{\bF}$ the set of all operations
$\otimes_{M_1,M_2,N_1,N_2,P_1,P_2}$ where $M_1$, $M_2$, $N_1$, $N_2$,
$P_1$ and $P_2$ are respectively $k_1\times \ell_1$, $k_2\times
\ell_2$, $k_1\times m_1$, $k_2\times m_2$, $\ell_1\times m_1$ and
$\ell_2\times m_2$-matrices and $k_1+k_2$, $\ell_1+\ell_2$ and
$m_1+m_2$ $\leq n$. Every term $t$ in $T(\cBR_n^{\bF}, \cBC_n^{\bF})$
defines, up to isomorphism, a $\bF^*$-graph denoted by
$val(t)$.

The operations in $\cBR_n^{\bF}$ can be defined in terms of disjoint
union and quantifier-free operations. Therefore, Theorem \ref{thm:5.1}
is still true if we replace $\cR_n^{(\bF,\sigma)}$ by $\cBR_n^{\bF}$.
The principal result of this section is the following.

\begin{thm} \label{thm:5.4} A $\bF^*$-graph has $\bF$-bi-rank-width at
  most $n$ if and only if it is isomorphic to some term $t$ in
  $T(\cBR_n^{\bF}, \cBC_n^{\bF})$. 
\end{thm}

The proof is similar to the one of Theorem \ref{thm:5.2}. The
following lemma is straightforward to verify.

\begin{lem}\label{lem:5.4} If $K=G\otimes_{M_1,M_2,N_1,N_2,P_1,P2} H$, then
  \begin{align*}
    \matind{\mat{K}}{V_G}{V_H}& = \Gamma_G^+\cdot M_1 \cdot
    (\Gamma_H^-)^T, & \matind{\mat{K}}{V_H}{V_G} &= \left(\Gamma_G^-\cdot
    M_2\cdot (\Gamma_H^+)^T\right)^T,\\ \Gamma_K^+ &= \begin{pmatrix}
      \Gamma_G^+\cdot N_1 \\ \Gamma_H^+\cdot P_1 \end{pmatrix}, &
    \Gamma_K^- &= \begin{pmatrix} \Gamma_G^-\cdot N_2 \\ \Gamma_H^-
      \cdot P_2 \end{pmatrix}.
  \end{align*}
  Moreover, $K$ is isomorphic to $H\otimes_{(M_2)^T,(M_1)^T,P_1,P_2,N_1,N_2} G$.
\end{lem}

\begin{lem} \label{lem:5.5} Let $t=c\bullet t'$ where $t'\in
  T(\cBR_n^{\bF},\cBC_n^{\bF})$ and $c\in
  Cxt(\cBR_n^{\bF},\cBC_n^{\bF})\backslash Id$. If $G=val(t)$ and
  $H=val(t')$, then $\matgind{V_H}{V_G\backslash V_H} =
  \Gamma_H^+\cdot B_1$ and $\matgind{V_G\backslash V_H}{V_H} =
  (\Gamma_H^-\cdot B_2)^T$ for some matrices $B_1$ and $B_2$.
\end{lem}

\begin{pf*}{Proof.} We prove it by induction on the structure of
    $c$, by showing in addition that $\Gamma_{G[V_H]}^+ =
  \Gamma_H^+\cdot C_1$ and $\Gamma_{G[V_H]}^- = \Gamma_H^-\cdot C_2$
  for some matrices $C_1$ and $C_2$. We identify two cases (the two
  other cases are similar by symmetry and Lemma \ref{lem:5.4}).

  \begin{description}
  \item[Case 1] $c=Id\otimes_{M_1,M_2,N_1,N_2,P_1,P_2} t''$.  We let $K=val(t'')$. Then
    $G=H\otimes_{M_1,M_2,N_1,N_2,P_1,P_2} K$. By Lemma \ref{lem:5.4},
    \begin{align*}
      \matgind{V_H}{V_G\backslash V_H}& = \Gamma_H^+\cdot M_1 \cdot
      (\Gamma_K^-)^T, & \matgind{V_G\backslash V_H}{V_H} &=
      \left(\Gamma_H^-\cdot M_2\cdot (\Gamma_K^+)^T\right)^T,\\ \Gamma_{G[V_H]}^+
      &= \Gamma_H^+\cdot N_1, & \Gamma_{G[V_H]}^- &= \Gamma_H^-\cdot
      N_2.
    \end{align*}
    We let $B_1=M_1 \cdot (\Gamma_K^-)^T$, $B_2=M_2\cdot
    (\Gamma_K^+)^T$, $C_1=N_1$ and $C_2=N_2$.
    
  \item[Case 2] $c=c'\otimes_{M,M',N,P} t''$ where $c'\in
    Cxt(\cBR_n^{\bF},\cBC_n^{\bF})\backslash Id$.  We let
    $K=val(t'')$ and $G'=val(c'\bullet t')$. Hence
    $G=G'\otimes_{M_1,M_2,N_1,N_2,P_1,P_2} K$. By Lemma
    \ref{lem:5.4}, 
    \begin{align*}
      \matgind{V_H}{V_G\backslash V_H} &= \begin{pmatrix}
        \matind{\mat{G'}}{V_H}{V_{G'}\backslash V_H} &&&
        \Gamma_{G'[V_H]}^+\cdot M_1 \cdot
        (\Gamma_K^-)^T\end{pmatrix},\\
        \matgind{V_G\backslash V_H}{V_H} &= \begin{pmatrix}
          \matind{\mat{G'}}{V_{G'}\backslash V_H}{V_H} &&&
          \left(\Gamma_{G'[V_H]}^-\cdot M_2 \cdot (\Gamma_K^+)^T\right)^T\end{pmatrix}
    \end{align*}
    By inductive hypothesis,
    $\matind{\mat{G'}}{V_H}{V_{G'}\backslash V_H} = \Gamma_H^+\cdot
    B_1'$ and $\matind{\mat{G'}}{V_{G'}\backslash V_H}{V_H}=
    (\Gamma_H^-\cdot B_2')^T$. Moreover, $\Gamma_{G'[V_H]}^+=
    \Gamma_H^+\cdot C_1'$ and $\Gamma_{G'[V_H]}^-=
    \Gamma_H^-\cdot C_2'$.  Therefore, letting
    \begin{align*}
      B_1 &= \begin{pmatrix} B_1' && C_1'\cdot M_1\cdot
        (\Gamma_K^-)^T \end{pmatrix}, & B_2 &= \begin{pmatrix} B_2'
        && C_2'\cdot M_2\cdot (\Gamma_K^+)^T \end{pmatrix}, \\ C_1 &=
      C_1'\cdot N_1, & C_2 &=C_2'\cdot N_2
    \end{align*}
    concludes the proof. \qed
  \end{description}
\end{pf*}

The following proves the ``if direction'' of Theorem \ref{thm:5.4}.
 
\begin{prop}\label{prop:5.3} If $G$ is  isomorphic
    to $val(t)$ for a term $t$ in $T(\cBR_n^{\bF},
    \cBC_n^{\bF})$, then $\fbrwd{G} \leq n$.
\end{prop}

\begin{pf*}{Proof.} Let $T$ be the syntactic tree of $t$. By
  definition, there exists a bijective function $\cL:V_G\to L_T$ where
  $L_T$ is the set of leaves of $t$, then of $T$. We let $(T,\cL)$ be
  a layout of $V_G$. In order to prove that the $\fbicutrk_G$-width of
  $(T,\cL)$ is at most $n$, it is sufficient to prove that for each
  subgraph $H$ of $G$ associated to a sub-term $t'$ of $t$,
  $\fbicutrk_G(V_H)\leq n$. However, we have proved in Lemma
  \ref{lem:5.5} that $\matgind{V_H}{V_G\backslash V_H} =
  \Gamma_H^+\cdot B_1$ and $\matgind{V_G\backslash V_H}{V_H} =
  \left(\Gamma_H^-\cdot B_2\right)^T$ for some matrices $B_1$ and
  $B_2$. And since each such $H$ is $\bF^{k_1,k_2}$-colored where
  $k_1+k_2\leq n$, we are done.\qed
\end{pf*}

The following proves the ``only if direction'' of Theorem
\ref{thm:5.4}. 

\begin{prop}\label{prop:5.4} If $\fbrwd{G}\leq n$, then $G$ is
  isomorphic to $val(t)$ for a term $t$ in
  $T(\cBR_n^{\bF},\cBC_n^{\bF})$.
\end{prop}

\begin{pf*}{Proof.} Assume first that $G$ is connected. Let $(T,\cL)$
  be a layout of $V_G$ of $\fbicutrk_G$-width at most $n$. We pick an
  edge of $T$, subdivide it and root the new tree $T'$ by considering
  the new node as the root. For each node $u$ of $T'$, we let $G_u$ be
  the subgraph of $G$ induced by the vertices that are in
  correspondence with the leaves of the sub-tree of $T'$ rooted at
  $u$. We let $r_1(u)$ be $\matgind{V_{G_u}}{V_G\backslash V_{G_u}}$
  and $r_2(u)$ be $\matgind{V_G\backslash V_{G_u}}{V_{G_u}}$.

  \begin{lem}\label{lem:5.6} For each node $u$ of $T'$, we can
    construct a term $t_u$ in $T(\cBR_n^{\bF},\cBC_n^{\bF})$ such that
    $val(t_u)$ is isomorphic to $G_u$ and is a
    $\bF^{r_1(u),r_2(u)}$-bi colored graph. There exists moreover
    vertex-bases $X_u^+$ and $X_u^-$ of, respectively,
    $\matgind{V_{G_u}}{V_G\backslash V_{G_u}}$ and
    $\left(\matgind{V_G\backslash V_{G_u}}{V_{G_u}}\right)^T$ such
    that $\matgind{V_{G_u}}{V_G\backslash V_{G_u}} =
    \Gamma_{val(t_u)}^+\cdot \matgind{X_u^+}{V_G\backslash V_{G_u}}$
    and $\matgind{V_G\backslash V_{G_u}}{V_{G_u}} =
    \matgind{V_G\backslash V_{G_u}}{X_u^-}\cdot
    (\Gamma_{val(t_u)}^-)^T$.
  \end{lem}

  It is clear that if $r$ is the root of $T'$, then $G=val(t_r)$ where
  $t_r$ is the term in $T(\cBR_n^{\bF},\cBC_n^{\bF})$
  constructed in Lemma \ref{lem:5.6}.

  Assume now that $G$ is not connected and let $G_1,\ldots, G_m$ be
  the connected components of $G$. By Lemma \ref{lem:5.6}, we can
  construct terms $t_1,\ldots, t_m$ that defines, up to
  isomorphism, respectively, $G_1,\ldots, G_m$. It is clear
  that $((\ldots((t_1\otimes_{O,O,O,O,O,O} t_2) \otimes_{O,O,O,O,O,O}
  t_3)\ldots) \otimes_{O,O,O,O,O,O} t_m)$ is isomorphic to $G$ where
  $O$ is the null matrix of order $1$. This concludes the proof of the
  proposition and therefore of Theorem \ref{thm:5.4}. \qed

  \begin{pf*}{Proof of Lemma \ref{lem:5.6}.} We prove it by
    induction on the number of vertices of $G_u$. Let $u$ be a node in
    $T'$. 

    If $G_u$ is a single vertex $x$, then since $G$ is connected we
    let $t_u:= \const{1\cdot 1}$ and $X_u^+:=X_u^-:=\{x\}$. It is clear
    that $t_u$, $X_u^+$ and $X_u^-$ verify the statements of the lemma.

    Assume now that $G_u$ has at least two vertices. Then $u$ has two
    sons $u_1$ and $u_2$ so that $G_{u_1}$ and $G_{u_2}$ have less
    vertices than $G_u$. By inductive hypothesis, there exist
    $t_{u_i}$, $X_{u_i}^+$ and $X_{u_i}^-$, for $i=1,2$, verifying the
    statements of the lemma.  We let $r_1(u_1):=h$, $r_2(u_1)=h'$,
    $r_1(u_2)=k$ and $r_2(u_2):=k'$. We let $X_{u_1}^+ :=
    \{x_{i_1},\ldots, x_{i_h}\}$, $X_{u_1}^- := \{x_{j_1},\ldots,
    x_{j_{h'}}\}$, $X_{u_2}^+ := \{y_{t_1}, \ldots, y_{t_k}\}$ and
    $X_{u_2}^- := \{y_{s_1}, \ldots, y_{s_{k'}}\}$. We let $M_1:=
    \matgind{X_{u_1}^+}{X_{u_2}^-}$ and $M_2:=
    (\matgind{X_{u_2}^+}{X_{u_1}^-})^T$, and $H=val(t_{u_1})$ and
    $K=val(t_{u_2})$.

    \begin{claim}\label{claim:5.2} $\matgind{V_{G_{u_1}}}{V_{G_{u_2}}} =
      \Gamma_{H}^+ \cdot M_1 \cdot (\Gamma_{K}^-)^T$ and
      $\matgind{V_{G_{u_2}}}{V_{G_{u_1}}} = \left(\Gamma_{H}^- \cdot
      M_2 \cdot (\Gamma_{K}^+)^T\right)^T$.
    \end{claim}

    \begin{pf*}{Proof of Claim \ref{claim:5.2}.} Let $x\in
      V_{G_{u_1}}$ and $y\in V_{G_{u_2}}$. By inductive hypothesis, 
      \begin{align*}
        \matgind{x}{V_G\backslash V_{G_{u_1}}} &= \gamma_{H}^+(x)\cdot
        \matgind{X_{u_1}^+}{V_G\backslash V_{G_{u_1}}}, \\
        \matgind{V_G\backslash V_{G_{u_1}}}{x} &=
        \matgind{V_G\backslash V_{G_{u_1}}}{X_{u_1}^-}\cdot
        (\gamma_H^-(x))^T \\ \matgind{y}{V_G\backslash V_{G_{u_2}}} &=
        \gamma_{K}^+(y)\cdot \matgind{X_{u_2}^+}{V_G\backslash
          V_{G_{u_2}}},\\ \matgind{V_G\backslash V_{G_{u_2}}}{y} &=
        \matgind{V_G\backslash V_{G_{u_2}}}{X_{u_2}^-}\cdot
        (\gamma_K^-(y))^T.\\ \intertext{Hence,}
        \gamma_{H}^+(x)\cdot M_1\cdot (\gamma_{K}^-(y))^T
        &= \matgind{x}{X_{u_2}^-} \cdot
        (\gamma_{K}^-(y))^T = \matgind{x}{y},\\ \intertext{and}
        \gamma_H^-(x)\cdot M_2 \cdot (\gamma_K^+(y))^T & =
        \gamma_H^-(x)\cdot  (\matgind{X_{u_2}^+}{X_{u_1}^-})^T \cdot
        (\gamma_K^+(y))^T\\ &= (\matgind{X_{u_2}^+}{X_{u_1}^-}\cdot
        (\gamma_H^-(x))^T)^T \cdot (\gamma_K^+(y))^T \\ &=
        (\matgind{X_{u_2}^+}{x})^T \cdot (\gamma_K^+(y))^T \\ &=
        \left(\gamma_K^+(y)\cdot \matgind{X_{u_2}^+}{x} \right)^T =
         \matgind{y}{x}. \qed
      \end{align*}
    \end{pf*}

    It remains now to find vertex-bases $X_u^+$ and $X_u^-$ of,
    respectively, $\matgind{V_{G_u}}{V_G\backslash V_{G_u}}$ and
    $(\matgind{V_G\backslash V_{G_u}}{V_{G_u}})^T$, and matrices
    $N_1$, $N_2$, $P_1$ and $P_2$ such that
    $\matgind{V_{G_u}}{V_G\backslash V_{G_u}} =
    \left(\begin{smallmatrix} \Gamma_H^+\cdot N_1\\ \Gamma_K^+\cdot
      P_1\end{smallmatrix} \right) \cdot \matgind{X_u^+}{V_G\backslash
        V_{G_u}}$ and $\matgind{V_G\backslash V_{G_u}}{V_{G_u}} =
      \matgind{V_G\backslash V_{G_u}}{X_u^-} \cdot
      \left(\begin{smallmatrix} \Gamma_H^-\cdot N_2\\ \Gamma_K^-\cdot
        P_2\end{smallmatrix} \right)^T$.

      It is straightforward to verify that
      $\{\matgind{z}{V_G\backslash V_{G_u}}\mid z\in X_{u_1}^+\cup
      X_{u_2}^+\}$ generates the row space of
      $\matgind{V_{G_u}}{V_G\backslash V_{G_u}}$. Similarly,
      $\{(\matgind{V_G\backslash V_{G_u}}{z})^T\mid z\in X_{u_1}^-\cup
      X_{u_2}^-\}$ generates the row space of $(\matgind{V_G\backslash
        V_{G_u}}{V_{G_u}})^T$. Therefore, we can find vertex-bases
      $X_u^+\subseteq X_{u_1}^+\cup X_{u_2}^+$ and $X_u^-\subseteq
      X_{u_1}^-\cup X_{u_2}^-$ of, respectively,
      $\matgind{V_{G_u}}{V_G\backslash V_{G_u}}$ and
      $(\matgind{V_G\backslash V_{G_u}}{V_{G_u}})^T$. That means, for
      each $z\in X_{u_1}^+\cup X_{u_2}^+$, there exists a row vector
      $b_z$ such that $\matgind{z}{V_G\backslash V_{G_u}} = b_z\cdot
      \matgind{X_u^+}{V_G\backslash V_{G_u}}$. Similarly, for each
      $z'\in X_{u_1}^-\cup X_{u_2}^-$, there exists a row vector
      $b'_z$ such that $\matgind{V_G\backslash V_{G_u}}{z} = b'_z\cdot
      \matgind{V_G\backslash V_{G_u}}{X_u^-}$. We let $t_u = t_{u_1}
      \otimes_{M_1,M_2,N_1,N_2,P_1,P_2} t_{u_2}$ where:
      \begin{align*}
        N_1 &:= \begin{pmatrix} b_{x_{i_1}} & \cdots & b_{x_{i_h}} \end{pmatrix}^T & P_1
        &:= \begin{pmatrix} b_{y_{t_1}} & \cdots &
          b_{y_{t_h}} \end{pmatrix} ^T\\
        N_2 &:= \begin{pmatrix} b'_{x_{j_1}} & \cdots & b'_{x_{j_{h'}}} \end{pmatrix}^T & P_2
        &:= \begin{pmatrix} b'_{y_{s_1}} & \cdots & b'_{y_{s_{k'}}} \end{pmatrix} ^T
      \end{align*}
      From Claim \ref{claim:5.2} it remains to show that
      $\Gamma_H^+\cdot N_1 \cdot \matgind{X_u^+}{V_G\backslash
        V_{G_u}} = \matgind{V_{G_{u_1}}}{V_G\backslash V_{G_u}}$ and
      $\Gamma_K^+\cdot P_1 \cdot \matgind{X_u^+}{V_G\backslash
        V_{G_u}} = \matgind{V_{G_{u_2}}}{V_G\backslash V_{G_u}}$, and
      $ \matgind{V_G\backslash V_{G_u}}{X_u^-}\cdot (\Gamma_H^-\cdot
      N_2 )^T = \matgind{V_G\backslash V_{G_u}}{V_{G_{u_1}}}$ and $
      \matgind{V_G\backslash V_{G_u}}{X_u^-}\cdot (\Gamma_K^-\cdot P_2
      )^T = \matgind{V_G\backslash V_{G_u}}{V_{G_{u_2}}}$.  But, this
      is a straightforward computation by the definitions of $N_1$,
      $N_2$, $P_1$ and $P_2$, and $X_u^+$ and $X_u^-$. This concludes the
      proof of the lemma. \qed
  \end{pf*}
\end{pf*}

\section{Conclusion}\label{sec:6}

We extended the rank-width and some related results from the undirected graphs to the $C$-graphs. Presented results imply in particular that every MSOL-definable property can be checked in polynomial time on $C$-graphs, when $C$ is finite. Every open question for the undirected case are of course still relevant for the $C$-graphs.

Recently, some authors investigated the clique-width of multigraphs~\cite{COU10b} or weighted graphs~\cite{FL10}. These graphs can be seen as $\mathbb{N}$-graphs. It is straightforward to verify that the rank-width is not equivalent to the clique-width when $C$ is infinite. It would be interesting to investigate the rank-width over an infinite field, and in particular its algorithmic aspects: the recognition of $C$-graphs of bounded rank-width, and the property checking on $C$-graphs of bounded rank-width.

  \end{document}